\documentclass[aps,notitlepage,twocolumn,nofootinbib,floatfix,pra,10pt]{revtex4-2}

\usepackage{amsmath,amsfonts,amssymb,amsthm,bbm,graphicx,enumerate,times,mathtools,hyperref,physics,marvosym,wasysym}
\hypersetup{colorlinks,linkcolor=blue,citecolor=red,urlcolor=blue}
\usepackage{float}
\usepackage[ruled,longend]{algorithm2e}
\usepackage{fancyref}
\usepackage{etoolbox}
\usepackage{subcaption}
\usepackage{tabularx}
\usepackage{appendix,soul}
\usepackage[dvipsnames]{xcolor}
\usepackage[noabbrev,capitalise]{cleveref}

\begin{document}
\title{SpinQ: Compilation strategies for scalable spin-qubit architectures}

\author{N. Paraskevopoulos$^{1,2}$}

\author{F. Sebastiano$^{1,2}$}
\author{C. G. Almudever$^{3}$}
\author{S. Feld$^{1,2}$}

\affiliation{$^{1}$Quantum and Computer Engineering Department, Delft University of Technology, 2628 CD Delft, The Netherlands}
\affiliation{$^{2}$QuTech, Delft University of Technology,  2628 CJ Delft, The Netherlands}
\affiliation{$^{3}$Computer Engineering Department, Universitat Politècnica de València, Camino de Vera, s/n, 46022 València, Spain}


\begin{abstract}
Despite NISQ devices being severely constrained, hardware- and algorithm-aware quantum circuit mapping techniques have been developed to enable successful algorithm executions. Not so much attention has been paid to mapping and compilation implementations for spin-qubit quantum processors due to the scarce availability of experimental devices and their small sizes. However, based on their high scalability potential and their rapid progress it is timely to start exploring solutions on such devices. In this work, we discuss the unique mapping challenges of a scalable crossbar architecture with shared control and introduce \textit{SpinQ}, the first native compilation framework for scalable spin-qubit architectures. At the core of \textit{SpinQ} is the \textit{Integrated Strategy} that addresses the unique operational constraints of the crossbar while considering compilation scalability and obtaining a \textit{O(n)} computational complexity. To evaluate the performance of \textit{SpinQ} on this novel architecture, we compiled a broad set of well-defined quantum circuits and performed an in-depth analysis based on multiple metrics such as gate overhead, depth overhead, and estimated success probability, which in turn allowed us to create unique mapping and architectural insights. Finally, we propose novel mapping techniques that could increase algorithm success rates on this architecture and potentially inspire further research on quantum circuit mapping for other scalable spin-qubit architectures.
\end{abstract}

\maketitle

\section{Introduction}
The prospect of quantum computing advantage is steadily becoming a reality \cite{arute2019quantum,madsen2022quantum,huang2022quantum}. The community is anticipating further advances that will allow quantum computing systems to become practical and to reach computational advantage \cite{bravyi2022future}. With such advancements, quantum computing systems are expected to solve a plethora of classically intractable problems. Until then, current quantum systems belong to the so-called Noisy Intermediate-Scale Quantum (NISQ) era \cite{preskill2018quantum}, in which devices can only handle small-sized quantum circuits. This is due to limitations in increasing the number of qubits and high operational errors with the latter causing rapid quantum information deterioration. Combined with more hardware constraints, such as cross-talk and limited classical-control resources \cite{ladd2010quantum,almudever2017engineering}, successful quantum circuit execution is a difficult feat. Scientists, both in academia and industry, face major engineering challenges in building both hardware and corresponding system software.

During the NISQ era, there have been significant efforts \cite{lao2021timing,murali2019noise,lao20222qan,nishio2020extracting,bandic2022full,murali2019full,quetschlich2022predicting,9736599,tannu2019not,pozzi2020using,zulehner2018efficient} to extract the most out of these resource-constrained and error-prone quantum computing systems. One of the approaches to do so is by developing hardware- and algorithm-aware quantum circuit mapping techniques to maximize performance. In general terms, mapping refers to the process of modifying potentially hardware-agnostic quantum circuits in such a way that they can be run on a given quantum computing device by respecting all of its constraints while optimizing performance (e.g., algorithm success rate). So far, several mapping techniques have been developed mostly for superconducting and ion-trap qubit devices, as they are nowadays one of the most well-recognized and most-developed qubit implementation technologies in terms of qubit counts and availability to users. However, spin-qubits emerge as a promising technology for scaling up quantum computing systems mainly due to their high integration potential \cite{RevModPhys.85.961,PhysRevA.57.120,vandersypen2017interfacing,veldhorst2015two,zajac2015reconfigurable,watson2018programmable}. Therefore, the scientific community is envisioning two-dimensional spin-qubit architectural proposals that could alleviate some of the major challenges towards scalability. Recently, a crossbar array \cite{borsoi2022shared} has been experimentally demonstrated, showing great promise for architectures with shared control. Such scalable architectural designs come with a new set of hardware constraints for which novel quantum circuit mapping techniques need to be developed.

In this paper, we present \textit{SpinQ}, the first native compilation framework focusing on scalable spin-qubit architectures. To this purpose, we target the so-called crossbar architecture proposed in \cite{li2018crossbar}. By creating a deep understanding of its operational constraints, we draw a clear picture of unique mapping challenges that arise in comparison to other qubit technologies. \textcolor{black}{We have devised a novel compilation approach called the \textit{Integrated Strategy}, a method inspired by mapping solutions found in \cite{morais2019mapping,helsen2018quantum}. Rather than seeking pure optimality, this strategy prioritizes scalability to harness the potential of scalable spin-qubit architectures. This pioneering compilation strategy uniquely and effectively navigates the rigid constraints of the crossbar architecture, doing so without adding to the computational complexity in comparison to alternative proposals. Yet, it's important to note that the current iteration does restrict the parallelization of certain gates, indicating room for improvement. However, this design has been created with future advancements in mind while keeping its time complexity in check. Our aim, through the elucidation of our results, is to highlight the imperative nature of comprehensive performance evaluation of emerging architectures and mapping techniques. Through our results, we aim to provide key insights into the importance of performing an extensive performance evaluation process of novel architectures and mapping techniques. With this compilation framework, we not only enable quantum algorithm executions on scalable spin qubit hardware but, perhaps more importantly, we form insights on the behavior and performance of this new breed of architectures. It also offers design guidelines vital for steering future breakthroughs in both hardware and software.}

The main contributions of this paper are:
\begin{enumerate}
    \item An in-depth analysis of mapping challenges in order to create novel mapping techniques for spin-qubit crossbar architectures. 
    \item \textit{SpinQ} -- The first native compilation framework dedicated to scalable spin-qubit architectures which utilizes a more scalable compilation strategy compared to previous proposals.
    \item A thorough performance analysis of the main sources of gate/depth overhead and estimated success probability when mapping well-defined quantum algorithms on the crossbar architecture.
    \item Deriving algorithmic- and hardware-specific mapping insights for the crossbar architecture and potentially other spin-qubit architectures from a scalability point of view.
\end{enumerate}

The remainder of this paper is structured as follows: In Sec. \ref{Spin qubits as a scalable platform}, the current progress and challenges of scalable spin-qubit architectures are presented. In Sec. \ref{The crossbar architecture}, the crossbar architecture is introduced as a potential candidate in scaling quantum devices in two dimensions, as well as its native operations. In Sec. \ref{Mapping challenges of a crossbar architecture}, we comprehensively analyze the unique challenges of mapping quantum algorithms on the crossbar architecture, which require novel mapping techniques. Then, in Sec. \ref{Overview of SpinQ}, we introduce \textit{SpinQ} -- the first native compilation framework for scalable spin-qubit architectures. In Sec \ref{Experimental Methodology} we refer to our experimental methodology and In Sec. \ref{Evaluation and analysis} we thoroughly analyze the performance of \textit{SpinQ} when mapping a broad and well-defined range of quantum algorithms on the crossbar architecture after which we form architectural and mapping insights. In Sec. \ref{Discussion and future directions}, we discuss potential improvements of our compilation strategy, and we compare its computational complexity to previous proposals. Finally, we conclude our work in Sec. \ref{Conclusion}.

\section{Spin qubits as a scalable platform} \label{Spin qubits as a scalable platform}

To fulfill the promise \cite{preskill2018quantum} of quantum computers being machines that solve some classically intractable problems, substantial system sizes have to be reached, i.e., a large number of qubits \cite{almudever2017engineering,gidney2021factor}. It still remains to be seen which qubit implementation technologies (e.g., superconducting, trapped ions, quantum dots, photonics, defect-based on nitrogen-vacancy diamond centers) will succeed in scaling up quantum computing systems with high-quality qubits \cite{resch2019quantum,chatterjee2021semiconductor}. Among them, spin qubits in quantum dots are a promising technology for scalable quantum computers due to the maturity of the semiconductor industry, the capability of \textcolor{black}{high integration on a single die compared to other qubit technologies  (the physical space of $1$ transmon qubit can fit $\sim1000$ spin qubits along with classical control electronics), long coherence times (close to 20$\mu$s), and the ability to operate in super-kelvin temperatures (up to 4 kelvin) \cite{yoneda2018quantum,camenzind2022hole,hendrickx2021qubit,chatterjee2021semiconductor,veldhorst2015two,RevModPhys.85.961,PhysRevA.57.120,vandersypen2017interfacing,veldhorst2015two,zajac2015reconfigurable,watson2018programmable}.} 

Despite the advantages just mentioned, there are still several challenges today towards scaling spin-qubit devices in a sustainable manner. One major challenge is the wiring scheme between the quantum processor and the classical interface, the so-called interconnect bottleneck \cite{vandersypen2017interfacing}. Formally, the interconnect bottleneck is described by Rent's exponent \cite{franke2019rent}, which is a measure of optimization in the wiring scheme in both classical and quantum processors. The existing scheme in most quantum devices, which has at least one control line per qubit, is not scalable in the long term. This is mostly due to the fact that dilution refrigerators have an upper limit to I/O cable capacity and that more cables will progressively make it harder to reach the desired milli-Kelvin temperature due to higher heat dissipation. Therefore, qubit architectures and classical-control electronics have to support multi-qubit shared-control that requires a sub-linear number of control lines alongside an increasing number of qubits. In other words, each control line needs to address multiple qubits to effectively mitigate the interconnect bottleneck when scaling up quantum hardware.

Going a step further, the inability to achieve a scalable wiring scheme also originates from the low device uniformity achieved by today's fabrication tools. In most cases, this implies that qubits can not be made homogeneous enough to control them effectively in a scalable architecture. The low uniformity results in resonance frequency deviations or other control variations. This means that in an inhomogeneous device, a driving signal for a particular operation will have to vary from one qubit to another to get the same outcome \cite{meyer2022electrical,vandersypen2017interfacing,li2018crossbar}. This makes it difficult to successfully control many qubits with the same line, thus contributing to the wiring scheme challenge (i.e., the interconnect bottleneck). 

There have been significant efforts \cite{li2018crossbar,boter2022spiderweb,vandersypen2017interfacing,hill2015surface,franke2019rent,paquelet2020multiplexed,pauka2019cryogenic,veldhorst2017silicon} to reduce the number of control lines reaching the qubits as devices become ever denser. Such efforts take advantage of the miniaturization capabilities of spin qubits and the large-scale integration of solid-state circuits to address the aforementioned challenges. However, current experimental work has primarily been focused on one-dimensional spin-qubit arrays of small sizes \cite{vandersypen2017interfacing}, which are not easily scalable. Recently, a $2 \times 2$ spin-qubit processor \cite{hendrickx2021four} and a $4 \times 4$ spin-qubit device based on a crossbar architecture \cite{borsoi2022shared,li2018crossbar} with shared control have demonstrated the potential to scale spin-qubit devices in two dimensions. Therefore, there will be a need, as the technology is advancing and further reducing Rent's exponent, to effectively map quantum algorithms on two-dimensional devices such as the crossbar architecture, which comes not only with limited qubit connectivity but also with a new set of constraints. This creates an opportunity to explore its mapping challenges and propose novel solutions.

However, mapping techniques are not studied as much as other qubit technologies such as superconducting and ion traps. In addition, the sample space of architectural proposals is sparse and lacks a detailed description of hardware constraints \textcolor{black}{\cite{boter2021spider,hill2015surface,veldhorst2017silicon}}. In combination with the lack of available devices for testing leads to a lack of a proper evaluation tool capable of benchmarking various quantum algorithms. Therefore, it also remains unclear whether existing techniques could be applicable. Then, even if such techniques are realized they could be incompatible with existing quantum compilation tools made for other qubit technologies. This could be due to completely different development requirements imposed by the particular spin-qubit constraints and their scalability prospects. In other words, a dedicated compilation framework for spin-qubit architectures with a focus on scalability is still missing. All these obstacles make it difficult to fully explore the possibilities and compare various architectural proposals under relevant application categories.

\section{The Crossbar Architecture} \label{The crossbar architecture}

\begin{figure}
    \centering    
    \includegraphics[width=0.5\textwidth]{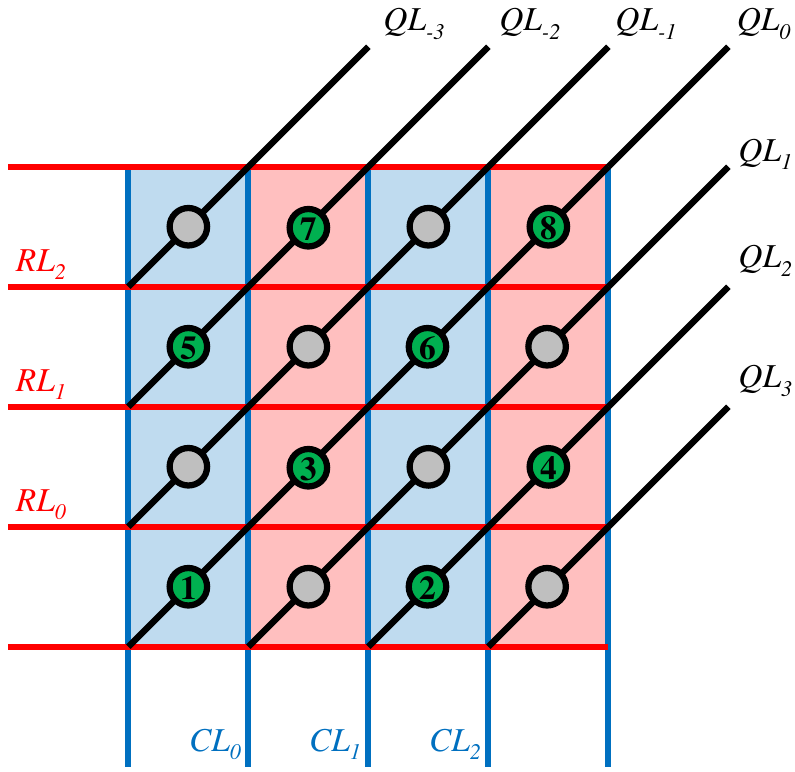}
    \caption{Schematic overview of the crossbar architecture and operational control lines \cite{li2018crossbar}.}
    \label{fig:crossbar}

\end{figure}

The crossbar architecture for arranging spin qubits was introduced in \cite{li2018crossbar} as a scalable solution to the interconnect bottleneck. Inspired by the crossbar architecture used in today’s classical processors \cite{boter2022spiderweb,li2018crossbar}, it adopts a similar characteristic, namely shared control. This achieves a quadratic reduction in control lines per qubit \cite{helsen2018quantum} and opens up the possibility for high integration of up to $1,000$ qubits in a single package. 

In this crossbar architecture, qubits are defined by electron spin states in Si-based quantum dots. \textcolor{black}{A Si-based quantum dot is a layer-structured semiconductor device that can confine a single electron with proper gate electrode control, after which its quantum-mechanical spin can define a physical qubit \cite{hanson2007spins}}. In Figure \ref{fig:crossbar}, we illustrate a schematic overview of the crossbar architecture in which every site (circles) represents a quantum dot, some of which are occupied by spin-qubits (numbered, green circles). Spin qubits are usually sparsely initialized in a pattern to reduce potential cross-talk and to allow for long-range entanglement through shuttling qubits across the array \cite{li2018crossbar}. In this case, a checker-board pattern provides these benefits. Finally, the crossbar architecture requires high fabrication uniformity of its materials to minimize operational errors. It is possible, however, to mitigate such errors or even vanish them by operating the crossbar at low magnetic fields and with proper tuning (e.g., separated resonance frequencies between columns). Furthermore, a crossbar module is envisioned to be self-contained and duplicated in a network of multiple crossbar modules. This can provide the means to realize quantum error correction (QEC) in large-scale systems enabled by fast-shuttling, low-error communication links.

It is now important to focus on the three different kinds of shared control lines used to perform operations on qubits: vertical (column line, CL), horizontal (row line, RL), and diagonal (qubit line, QL). Notably, each line affects all the sites that it is connected to. For instance, in Fig. \ref{fig:crossbar} line $QL_{-2}$ affects the sites in which spin-qubits $5$ and $7$ reside. This imposes some particular restrictions in the parallelization of instructions, which we will discuss in Sec. \ref{Mapping challenges of a crossbar architecture}. Below, we will abstractly describe the control properties for executing native gates of the crossbar architecture. \textcolor{black}{Note that any non-native gates can be decomposed into native ones as explained in Sec. \ref{Decomposition of quantum gates}.} A more detailed explanation is provided in \cite{helsen2018quantum}.

\subsection{Qubit shuttling} \label{Qubit shuttling}
In the crossbar architecture, qubits can be moved around by performing shuttling operations. In a shuttle operation, vertical or horizontal lines are used as barrier gates, depending on the direction. Lowering or raising these barriers can create pathways from which qubits can move (shuttle) orthogonally from one site to another with the use of DC signals through the diagonal lines. \textcolor{black}{Specifically, when a barrier, separating a qubit and an empty quantum dot, is lowered then it can move pushed/attracted by the voltage difference of the QL lines going through the origin and destination sites.} Fig. \ref{fig:shuttle_example} (a) shows an example of shuttling, in which qubit $3$ is moving one site to the left. \textcolor{black}{The order in which the control line signals should be pulsed and other requirements are analyzed in Sec. \ref{Parallelization of quantum operations}. Fig. \ref{fig:shuttle_example} (b) shows the potential parallel shuttle operations to shuttling qubit 3 left. In order to do that, the requirements of shuttling qubit 3 need to be compatible with the others and satisfied in the same order. Note that the larger the crossbar topology, the more difficult it gets to parallelize shuttle operations. A deeper analysis of parallelization possibilities and challenges is given In Sec. \ref{Mapping challenges of a crossbar architecture}.}  Although this architecture can support gate-based communication with two subsequent \(\sqrt{SWAP}\)s (see Sec. \ref{Two-qubit gates}), resulting in a $SWAP$ gate similar to superconducting qubits, shuttling qubits is preferred due to higher operation fidelity and shorter execution time. It should be noted that shuttling horizontally, i.e., between columns, causes a Z rotation (see Sec. \ref{Z-phase rotations}) which should be mitigated by timing well the next operation(s) \cite{li2018crossbar}.

\subsection{Two-qubit gates} \label{Two-qubit gates}
Two two-qubit gates are supported by the crossbar \textcolor{black}{architecture}, $CPHASE$ and \(\sqrt{SWAP}\), with the latter being chosen for this work due to its higher operational fidelity and faster execution time according to \cite{li2018crossbar}. \textcolor{black}{A \(\sqrt{SWAP}\) can be performed similarly to the requirements of a shuttle operation, analyzed in Sec. \ref{Qubit shuttling} and \ref{Parallelization of quantum operations}. The difference is that the two qubits need to be vertically adjacent (i.e., same column) and the fourth requirement related to the \textit{QL} lines going through the two sites needs to have equal voltages. Once these are satisfied, similar rules to shuttle parallelization are applied for parallel \(\sqrt{SWAP}\)s as explained in Sec. \ref{Qubit shuttling}. Finally, it is possible to parallelize two-qubit gates and shuttle operations as long as all their constraints are satisfied.}

\subsection{Z rotations} \label{Z-phase rotations}
In the crossbar architecture, single-qubit gate rotations should be separated into two categories: Z rotations and $X$ or $Y$ rotations. 

Z qubit rotations can be controlled by a well-timed qubit shuttling to and from a neighboring column \cite{li2018crossbar,helsen2018quantum,morais2019mapping}. \textcolor{black}{Due to the differences in Zeeman energies between the two column parities, timing is key. The imposing alternating magnetic fields on qubits rotate them in the Z axis and the longer they stay in the opposite column parity the more they rotate.} \textcolor{black}{Therefore, when the second shuttle is timed purposefully, the qubit can return to the initial position rotated at any angle. Besides this timing peculiarity between the two shuttles, parallelization constraints and mapping challenges are the same as qubit shuttling. Finally, it is possible to parallelize Z rotations, two-qubit gates, and shuttle operations in the same cycle when all requirements are satisfied.}

\subsection{X or Y rotations}
As for $X$ or $Y$ rotations, either all qubits belonging to red-colored columns or all qubits in blue-colored columns are rotated (see Fig. \ref{fig:crossbar}). This is called semi-global single-qubit rotation and is implemented by electron-spin-resonance \cite{veldhorst2014addressable,li2018crossbar}. \textcolor{black}{ A high-level representation of how a particular column parity is addressed is given in Fig. \ref{fig:scheme} (a). Depending on the duration of the applied magnetic field at the CL lines, the qubits can be rotated at any angle.}

\subsection{Measurement} \label{Measurement}
The readout process allows for local single qubit measurements by using the Pauli Spin Blockade (PSB) process \cite{fujita2017coherent}. With this process, the measurement outcome is determined by whether a qubit shuttle towards a horizontally adjacent ancilla qubit was successful or not. In this work, we considered an ideal measurement process in which no ancilla qubits are involved. 

\section{Quantum circuit mapping challenges of the crossbar architecture} \label{Mapping challenges of a crossbar architecture}

The mapping process of a quantum circuit plays an essential role in the successful execution of algorithms on a quantum computer. It consists of a cascade of routines that transform a (potentially hardware-agnostic) quantum circuit to a hardware-compatible version. However, current NISQ quantum processors are severely constrained and cannot run useful applications successfully yet, despite notable efforts in this field. 

Examples of hardware constraints are low qubit connectivity, cross-talk, reduced primitive gate set, low coherence time, fabrication imperfections, and limited classical-control resources. Therefore, a mapping process needs to consider such limitations and try to optimize performance as much as possible to increase the algorithm's success rate. So far, there are a plethora of proposed solutions which differ in strategy, methodology and performance metrics to optimize \cite{zulehner2018efficient,lao2021timing,murali2019noise,lao20222qan,nishio2020extracting,bandic2022full,murali2019full,quetschlich2022predicting,9736599,tannu2019not,pozzi2020using,li2019tackling}.

Such mapping techniques have been mostly developed for superconducting and ion-trap qubit devices. However, as of now, there is not much focus on spin-qubit architectures and their particular characteristics. Although spin-qubits are now in a rather early development stage, their scalability potential is undeniable, and therefore, it is timely to lay grounds for developing novel mapping techniques and inspire further research. As previously mentioned, in this work, we focus on the crossbar architecture that comes with a unique set of constraints that affect the parallelization of quantum operations, the application of $X$ or $Y$ rotations on individual qubits, and the routing of qubits (i.e., moving qubits around the topology).

\subsection{Parallelization of quantum operations} \label{Parallelization of quantum operations}

\begin{figure}[htpb]
    \centering
         \begin{subfigure}[b]{0.5\textwidth}
             \centering
             \includegraphics[width=\textwidth]{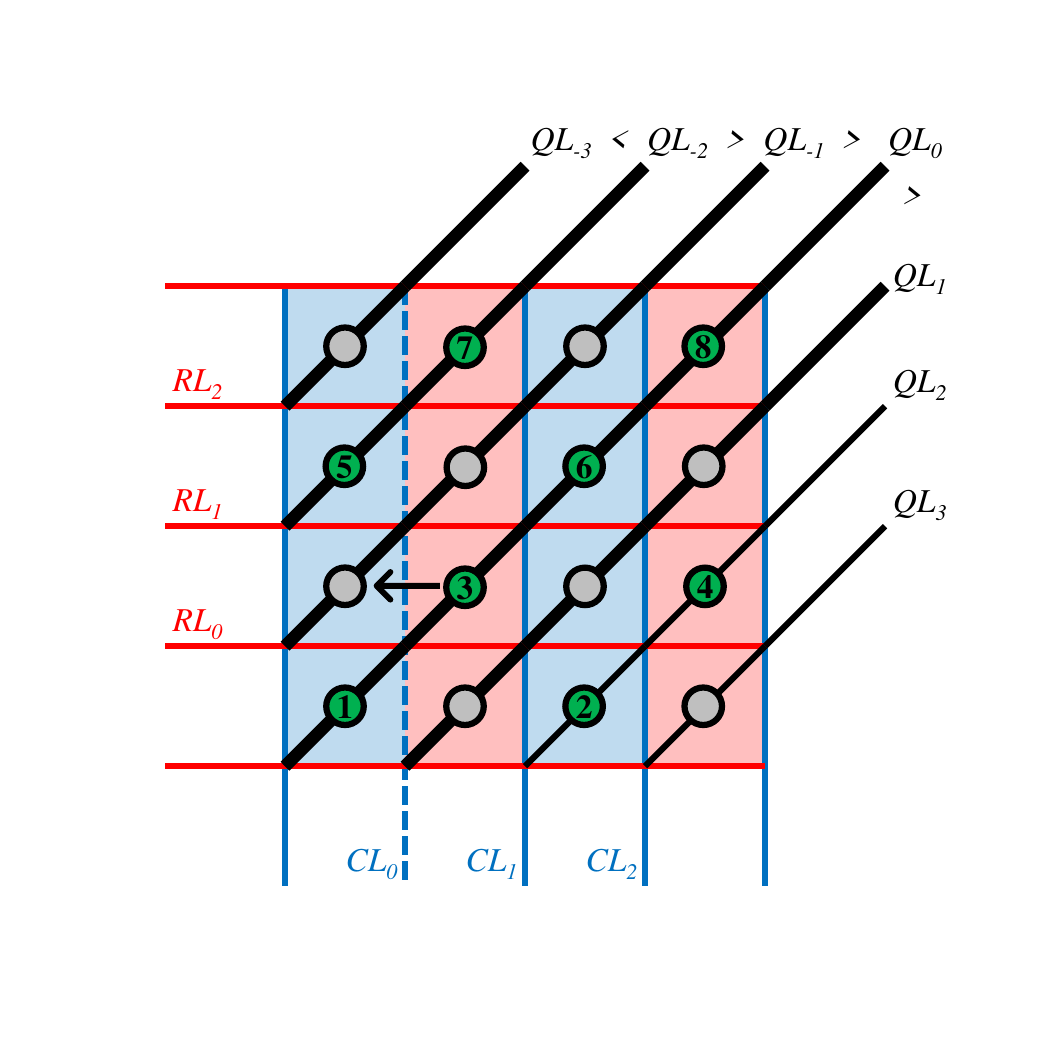}
             \caption{}
             \label{fig:shuttle_example}
         \end{subfigure}
        \hfill
         \begin{subfigure}[b]{0.5\textwidth}     
             \centering
             \includegraphics[width=\textwidth]{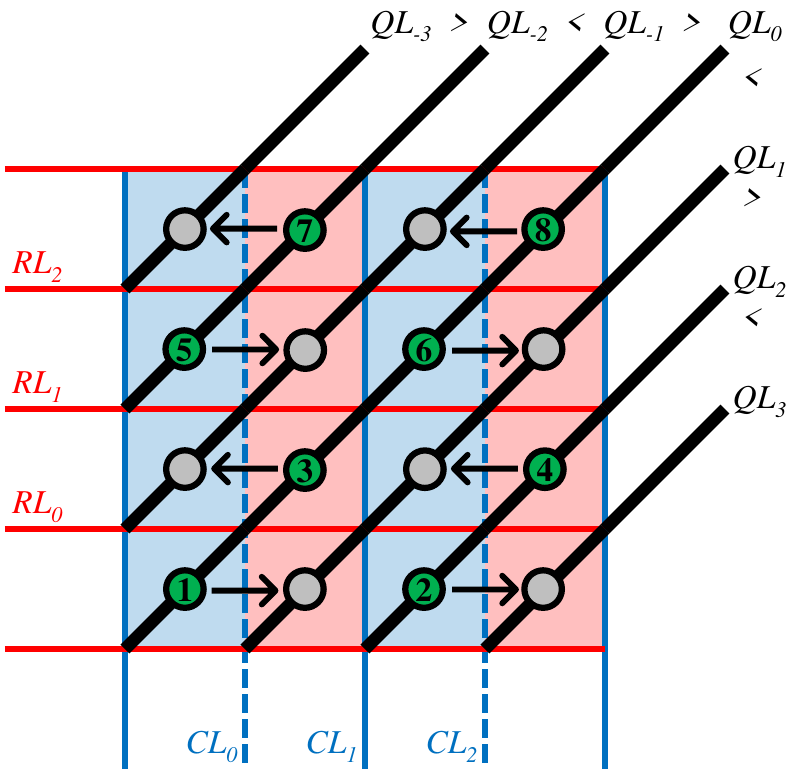}
             \caption{}
             \label{fig:shuttle_parall_example}
         \end{subfigure}
    \caption{\textcolor{black}{(a) Shuttling example of qubit 3 moving one site to the left. The barrier $CL_0$ between origin and destination site is lowered and voltage of $QL_{-1}$ is larger than $QL_{0}$. The order and signal requirements will be further explained in Sec. \ref{Parallelization of quantum operations} (b) Shows the potential of parallelizing shuttling operations that can be executed in parallel with shuttling left qubit 3. However, not all qubits can be shuttled arbitrarily at the same time due to the specific requirements or potential conflicts caused by mapping in this architecture. These challenges are analyzed in Sec. \ref{Mapping challenges of a crossbar architecture}.}}
    \label{fig:shuttle_example}
    
\end{figure}

\begin{figure}[htpb]
    \centering
    \includegraphics[width=0.5\textwidth]{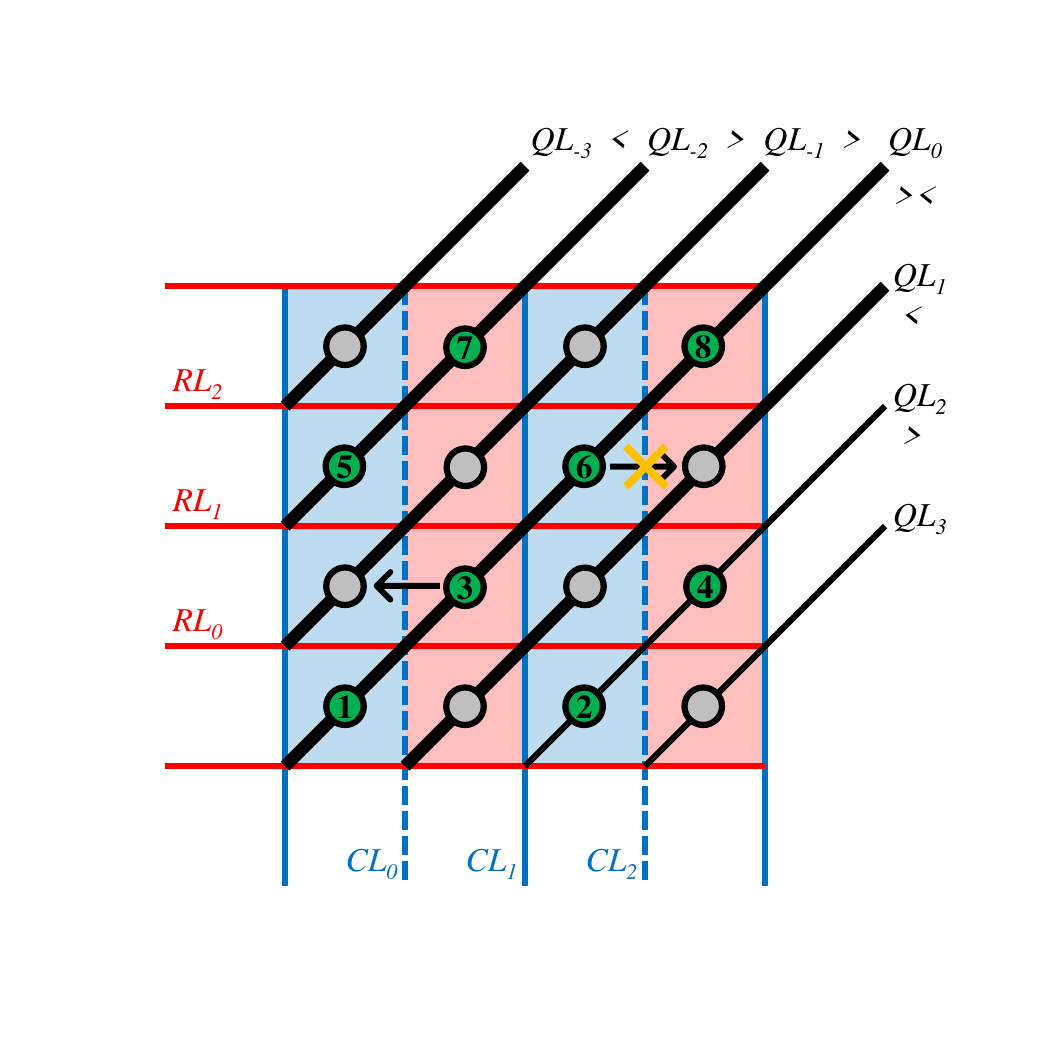}
    \caption{Parallelizing shuttles of qubit $3$ and $6$ is not allowed due to violation of constraints \textcolor{black}{shown as $QL_{0}><QL_{1}$.}}
    \label{fig:shuttle_example_conflict}
\end{figure}

Most of the operation parallelization restrictions of the crossbar architecture come from the fact that control lines are shared among multiple qubits and each line has a specific role and relation to one another. It should also be mentioned that most operations must be implemented with strict pulse durations and time intervals, depending on the addressed site \cite{li2018crossbar,helsen2018quantum}. Although such pulse durations have to be carefully considered in the mapping process by providing recent calibration data \cite{9736599,tannu2019not,nishio2020extracting}, in this work, we consider an ideal crossbar architecture, as such data are not available yet. Despite that, the mapping techniques proposed in this work are compatible with similar considerations and can be added once calibration data are available.

To better illustrate what conditions and constraints there are when trying to parallelize quantum operations, let us consider the following example in which two shuttles are tried in parallel. As shown in Fig. \ref{fig:shuttle_example} (a), the following requirements must be fulfilled \textcolor{black}{in that order} to shuttle qubit $3$ one site to the left:
\begin{enumerate}
    \item The destination site must not be occupied by another qubit.
    \item The barrier between destination and origin sites must be lowered. This is depicted as a dashed vertical $CL_0$ line.
    \item All barriers surrounding the origin and destination sites must be raised. This is shown as solid red RL ($RL_0$ and $RL_1$) and blue CL lines ($CL_1$ and the always-raised most-left CL line).
    \item The voltage going through the QL line of the destination site ($QL_{-1}$) must be higher than the one going through the origin site ($QL_0$). This is shown as $QL_{-1}>QL_{0}$ in the top-right of Fig. \ref{fig:shuttle_example} (a).
    \item To prevent other qubits in these two columns from shuttling, the voltage going through their QL lines must be higher than their adjacent empty sites. This is depicted as voltage level relations between QL lines. Note that QLs with no voltage relations are irrelevant for this particular shuttle operation.
\end{enumerate}

Now, we assume a shuttle of qubit $6$ to the right (as depicted in Fig. \ref{fig:shuttle_example_conflict}) in parallel to the left shuttle of qubit $3$. This implies that all previously listed requirements of qubit $3$ need to be satisfied along with the new ones of qubit $6$. However, the fourth requirement can not be satisfied as the $QL_{0}>QL_{1}$ relation we had before would have to be changed to $QL_{0}<QL_{1}$. If this change is allowed, we violate the fifth requirement of the first shuttle and, as a consequence, qubit $1$ will shuttle to the right. Therefore, we can not shuttle qubits $3$ and $6$ as such at the same time. \textcolor{black}{Contrary to that, in Fig \ref{fig:shuttle_example} (b) we were able to shuttle qubit 6 only because qubit 1 shuttles to the right at the same time. }

Thus, we see that scheduling parallel gates in the crossbar implies a strict simultaneous satisfaction of all signal requirements for each gate. \textcolor{black}{It also depends on the specific gate (operation) set to be parallelized and their corresponding qubit positions on the topology.} Any violation of the above conditions would potentially result in the shuttling of unwanted qubits, unwanted qubit interactions, or unknown qubit states. As seen in the previous example, performing quantum operations in parallel without affecting other qubits and meeting all signal requirements is not always possible regardless of qubit distance. In fact, it does not matter how far qubits are away from each other, but whether control lines are shared between them or not, and whether their operational requirements and relations match or not. \textcolor{black}{A collision is another conflict that could be caused by improper parallelization of shuttling gates that try to move two qubits toward the same site.} Unlike more popular qubit architectures based on superconducting or ion traps, this form of operational constraint is unique. On one hand, sharing control lines tackles the interconnect bottleneck, on the other hand, it intrinsically constraints its parallelization capabilities.

Finally, in other qubit architectures, it is possible to perform different gate types in parallel. In the crossbar architecture, this is not always the case. For example, applying single-qubit gates and shuttling operations at the same time is not possible, because, in the former, CL lines need to carry an alternating current (AC) signal (see Fig. \ref{fig:scheme_1}) while the latter require DC signals for raising or lowering the barriers.  

\subsection{\textcolor{black}{Mapping} of \texorpdfstring{$X$}{x} or \texorpdfstring{$Y$}{y} rotations on single qubits} \label{Application of X or Y rotations on single qubits}

\begin{figure*}
     \centering
     \begin{subfigure}[b]{0.495\textwidth}
         \centering
         \includegraphics[width=\textwidth]{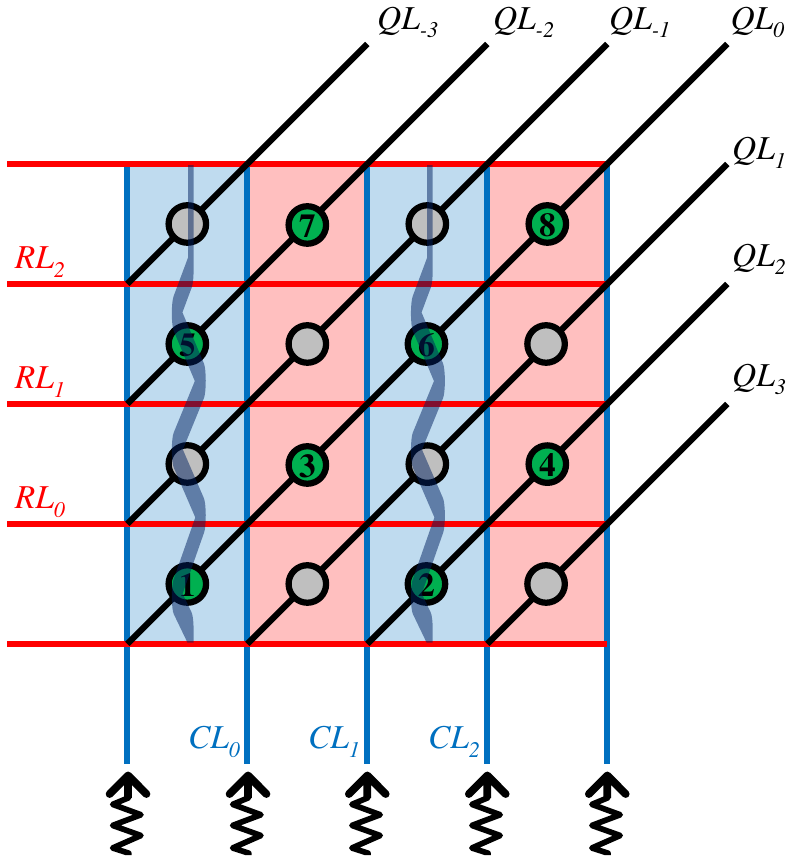}
         \caption{Step 1}
         \label{fig:scheme_1}
     \end{subfigure}
     \hfill
     \begin{subfigure}[b]{0.495\textwidth}
         \centering
         \includegraphics[width=\textwidth]{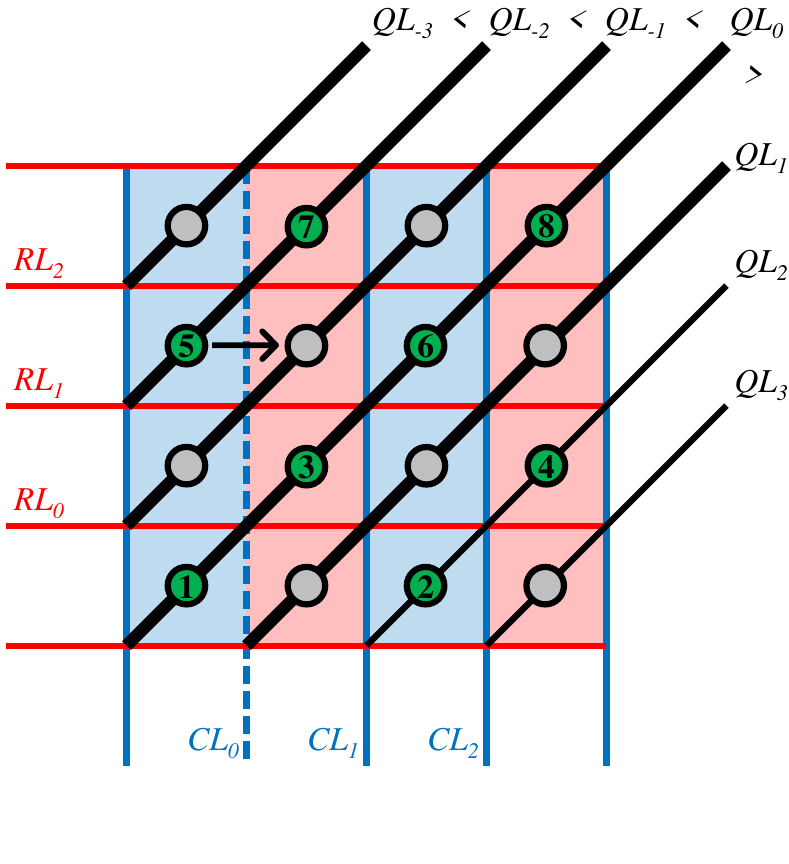}
         \caption{Step 2}
         \label{fig:scheme_2}
     \end{subfigure}
     \hfill
     \begin{subfigure}[b]{0.495\textwidth}
         \centering
         \includegraphics[width=\textwidth]{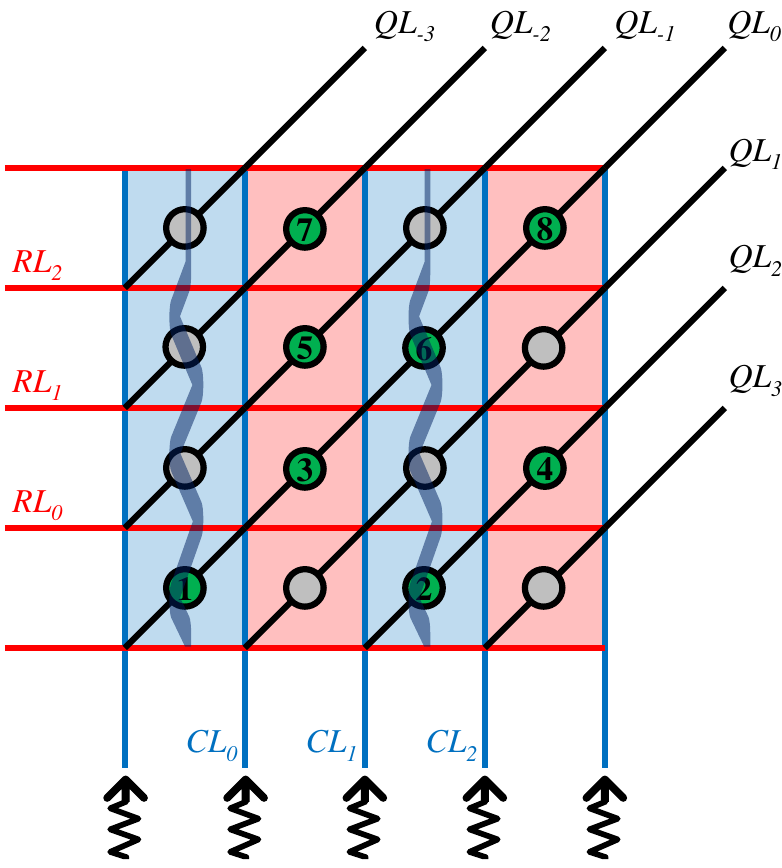}
         \caption{Step 3}
         \label{fig:scheme_3}
     \end{subfigure}
     \hfill
     \begin{subfigure}[b]{0.495\textwidth}
         \centering
         \includegraphics[width=\textwidth]{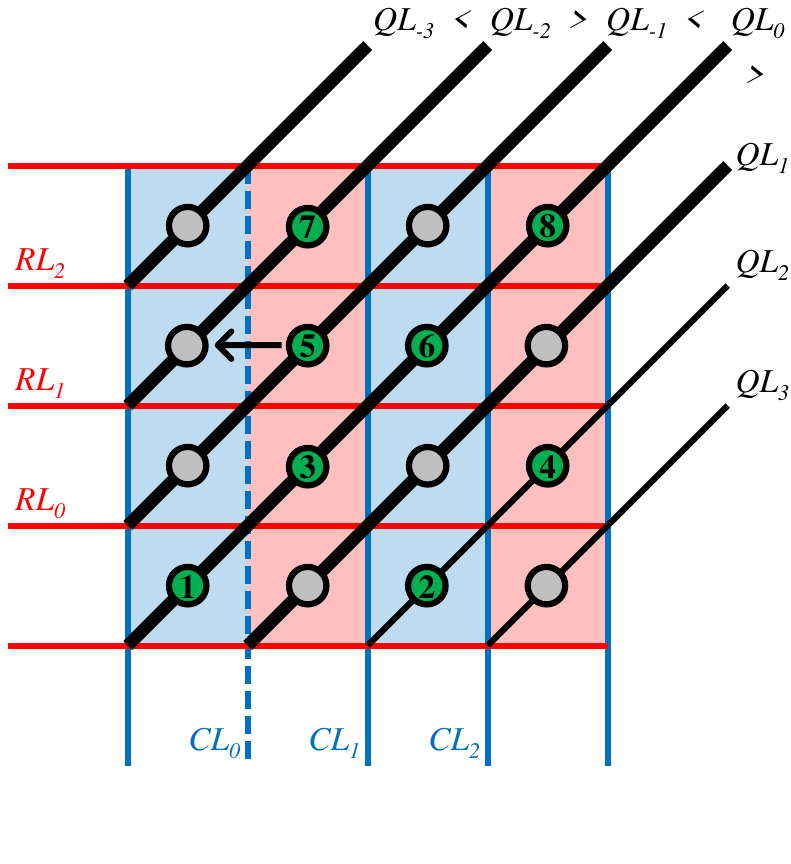}
         \caption{Step 4}
         \label{fig:scheme_4}
     \end{subfigure}
        \caption{Single-qubit gate on qubit 5: \textbf{(a) Step 1}: AC signals through the CL lines induce magnetic fields on qubits 1, 5, 6, and 2 belonging to the even column parity, thus rotating their state. The direction and frequency of these signals determine which columns (red or blue) and what rotation ($X$ or $Y$) will be applied to the corresponding qubits \cite{veldhorst2014addressable,li2018crossbar}. \textbf{(b) Step 2}: The targeted qubit 5 is moved with a shuttle operation to a different column parity. For this operation, $CL_0$ opens and closes as a barrier and the relevant diagonal lines (QL) create potential gradients to only allow for qubit 5 to move (shuttle). Note that in order to shuttle only qubit 5 all relevant QL lines need to have voltage relations with one another. \textbf{(c) Step 3}: An inverse rotation is applied again in all even columns containing qubits 1, 6, and 2, similarly to Step 1. \textbf{(d) Step 4}: Target qubit 5 is moved with a shuttle operation to the initial position.}
        \label{fig:scheme}         
\end{figure*}

As established in Sec. \ref{The crossbar architecture}, $X$ or $Y$ qubit rotations are implemented semi-globally, meaning that either all qubits in odd or even column parities will be rotated. However, during an arbitrary cycle, not all qubits in odd or even columns should be rotated. \textcolor{black}{Note that the notion of a cycle refers to the basic unit of time representing one step in a sequence of quantum gates applied to a set of qubits, and it may contain multiple gates.} Therefore, to compensate for unwanted $X$ or $Y$ rotations, one has to come up with a specific rotation mapping scheme such that only the targeted qubits are rotated. In this work, we have implemented the scheme introduced by \cite{helsen2018quantum}. We illustrate how it works in Fig. \ref{fig:scheme} in which we are interested in rotating only qubit $5$. This is another unique characteristic of this architecture as additional routing is needed to perform single-qubit rotations on specific qubits thus imposing new challenges to the mapping process.

\subsection{Routing of Qubits} \label{Routing Constraints}

\begin{figure}[htpb]
    \centering
    \includegraphics[width=0.5\textwidth]{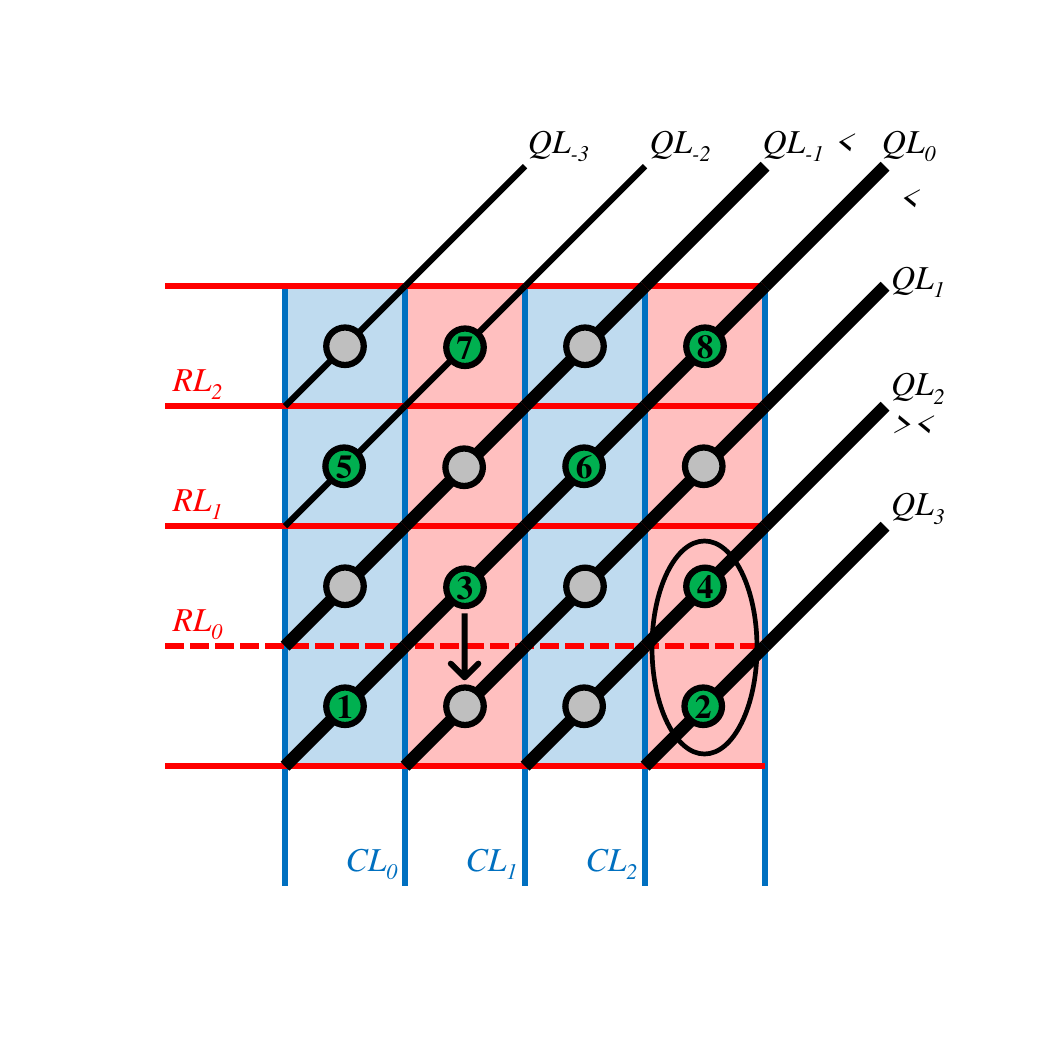}
    \caption{Example of a conflict: the operational requirements of shuttling qubit $3$ downwards have lowered the $RL_0$ barrier thus causing an unwanted interaction between qubits $2$ and $4$. \textcolor{black}{Additionally, $QL_2$ and $QL_3$ lines passing through these qubits need to satisfy the $5^{th}$ requirement described in Sec. \ref{Parallelization of quantum operations} and by doing so create a violation $QL_{2}><QL_{3}$. Note that $QL_1$ and $QL_2$ are signaled in this example but do not need to have a voltage relation requirement between them.}}
    \label{fig:Deadlock}
    
\end{figure}

We will now expand specifically on the qubit routing challenges:

Routing a qubit in the crossbar means that an electron (or a hole, depending on the materials) is physically "pushed" to an empty site (i.e., an empty quantum dot). This mechanism is similar to a quantum charged-coupled ion trap device (QCCD) where ions are shuttled through a common channel from trap to trap, assuming sufficient destination ion trap capacity \cite{bruzewicz2019trapped,murali2020architecting}. The QCCD architecture and the crossbar architecture fundamentally differ in topology, but both require special algorithms or additional routing routines to maintain control of qubit positions and avoid potential conflicts. \textcolor{black}{The topology of the crossbar is essentially a 2D square grid whereas a QCCD device resembles a bi-linear array with an "H" shape and in its corners ion traps are located -- each dedicated to a specific purpose during operation. The particular shape of a QCCD device creates different constraints for moving qubits or parallelizing operations compared to the crossbar architecture hence their mapping techniques are different even though both use shuttling.}

Shuttling qubits on the crossbar does not only depend on specific control signal requirements and available empty sites but on other qubit positions as well. We illustrate this fact with an example in Fig. \ref{fig:Deadlock}, in which a vertical shuttle operation of qubit $3$ is indicated by a black arrow. In this case, the horizontal barrier $RL_0$ has to be lowered and the QL lines have to be pulsed in certain voltage relations to allow for correct shuttling of only qubit 3. However, an unwanted interaction is caused between two other qubits in the same rows (qubits $2$ and $4$, circled), regardless of the $QL_2$ and $QL_3$ relation. Analogously, the same issue exists with a horizontal shuttle when having two horizontally adjacent qubits in the same columns where the shuttle takes place \cite{morais2019mapping,helsen2018quantum}. Lastly, there can be a blocked path conflict where there is no empty site for a qubit to shuttle to.

Therefore, a dedicated qubit routing algorithm for the crossbar architecture has to be developed to avoid collisions, blocked paths, and unwanted interactions. Furthermore, even if we had such a dedicated routing algorithm, the same conflicts have to be considered when rearranging gates in parallel during scheduling. For that, control signals and qubit positions must be carefully monitored within the mapping process. \textcolor{black}{We provide a summary of unique architectural features and operational constraints in Table \ref{fig:hardware_table} to clearly show these unique constraints.} From the description above, it is clear that both the routing and scheduling processes need to jointly work in a strategy to avoid conflicts and optimize for algorithm success rate. This will be the main characteristic of \textit{SpinQ}, presented in the following section.

\begin{table*}
    \centering
        \caption{\textcolor{black}{Summary of unique architectural features and operational constraints of the crossbar architecture.}}
\includegraphics[width=\textwidth]{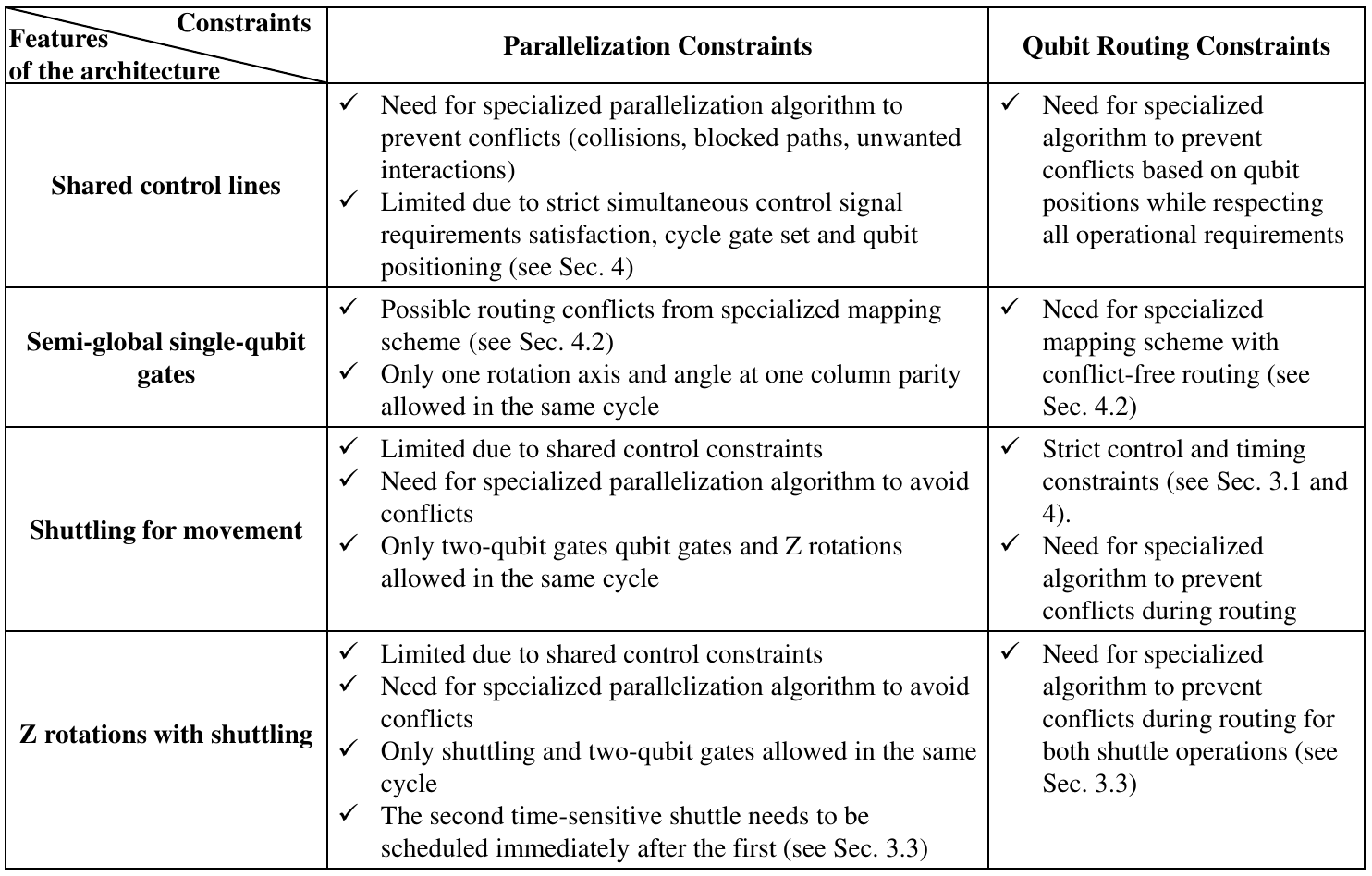}
    \label{fig:hardware_table}   
\end{table*}

\section{\textit{SpinQ} -- the first native compilation framework for scalable spin-qubit architectures} \label{Overview of SpinQ}

\begin{figure*}
    \centering
\includegraphics[width=0.8\textwidth]{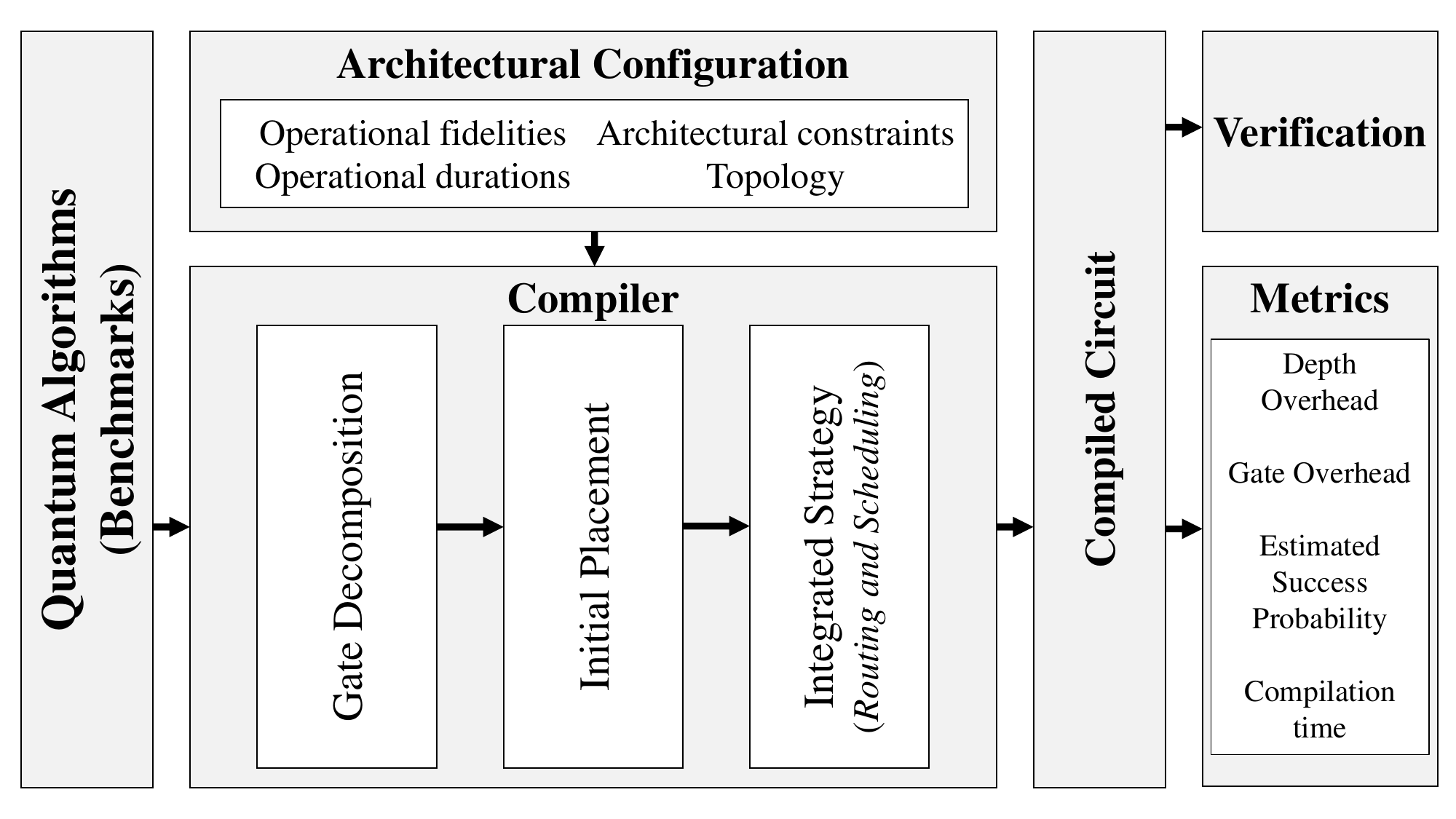}
    \caption{Overview of our SpinQ framework proposed in this paper.}
    \label{fig:Framework}    
\end{figure*}

In this work, we present the first native compilation framework -- \textit{SpinQ} -- dedicated to compiling and mapping quantum circuits onto scalable spin-qubit architectures, such as the previously described crossbar. We have based our mapping techniques on previous works from \cite{helsen2018quantum,morais2019mapping} while improving them from a scalability standpoint.

Fig. \ref{fig:Framework} shows the schematic structure of our framework. As \textbf{input}, \textit{SpinQ} accepts QASM format files that describe quantum circuits (used as benchmarks) in a device-independent manner. To increase the flexibility of our framework, particular characteristics of the crossbar architecture can be defined in an \textbf{architectural configuration} file. It can include custom operations and their particular attributes such as gate durations, mathematical description of the unitary matrices, associated gate fidelities, and architectural constraints, among others. Moving on, the \textbf{compiler} consists of a series of passes to decompose gates, route qubits, and schedule instructions. To address the unique mapping constraints of the crossbar architecture, we have conceptualized and developed the \textit{Integrated Strategy}. \textcolor{black}{The current implementation has room for improvement (see  Sec. \ref{Discussion and future directions}), however, our aim in this work is to study the behavior of algorithms in order to form deep insights about novel mapping techniques and spin-qubit architectures from a scalability perspective.} The compiler's output is a QASM file of the \textit{compiled circuit} which is compatible to be executed on the given architecture. Optionally, a \textbf{verification} step can take place to ensure the compiled circuit meets all operational constraints of the architecture without any conflicts. This step is implemented to be able to check the compatibility of architectural proposals that are not physically realized yet, such as the crossbar \cite{li2018crossbar} Finally, several \textbf{performance metrics} are extracted from the compiled circuit to evaluate algorithm performance.
In the next sections, we will further discuss each of the compiler elements.

\subsection{Compilation passes} \label{Compiler Stages}

The compiler consists of the following steps:

\subsubsection{\textbf{Decomposition of quantum gates}} \label{Decomposition of quantum gates}
Inputted QASM quantum circuits are first transformed into a custom-made intermediate representation (IR) data format. Quantum gates are then decomposed into gates native to the architecture based on the decomposition sequences specified in the architectural configuration file. \textcolor{black}{These sequences are reported in \cite{morais2019mapping}.}

\subsubsection{\textbf{Physical initialization of spin qubits}}
A checkerboard pattern has been proposed \cite{li2019tackling} to allow space for data and ancilla qubits to move \cite{helsen2018quantum,morais2019mapping}. The physical space achieved between the qubits not only facilitates shuttling that avoids possible conflicts but also reduces crosstalk and enables surface code error correction \cite{li2018crossbar}. As we will discuss later, maintaining this placement pattern throughout a circuit execution plays an integral role in the \textit{Integrated Strategy}. Having said that, initializing qubits in alternative patterns and changing them during execution is possible. This flexibility offered by the spin-qubit technology can be particularly advantageous to highly specialized mapping techniques for the crossbar as well as for other architectural proposals.

\subsubsection{\textbf{Virtual-to-physical qubit initial placement}}
The current version of \textit{SpinQ} associates virtual qubits of an algorithm with physical qubits in a one-to-one manner by numbering the physical qubits from left to right and from bottom to top as shown in Fig. \ref{fig:crossbar}. In the results sections \ref{Evaluation and analysis} and \ref{Discussion and future directions}, we will provide insights on how common initial placement algorithms can be adapted to improve the performance of spin-qubit architectures, such as the crossbar.

\subsubsection{\textbf{Integrated Strategy for Routing and Scheduling}} \label{Integrated Strategy for Routing and Scheduling}

\begin{algorithm*}[htpb]
    \centering
\includegraphics[width=0.9\textwidth]{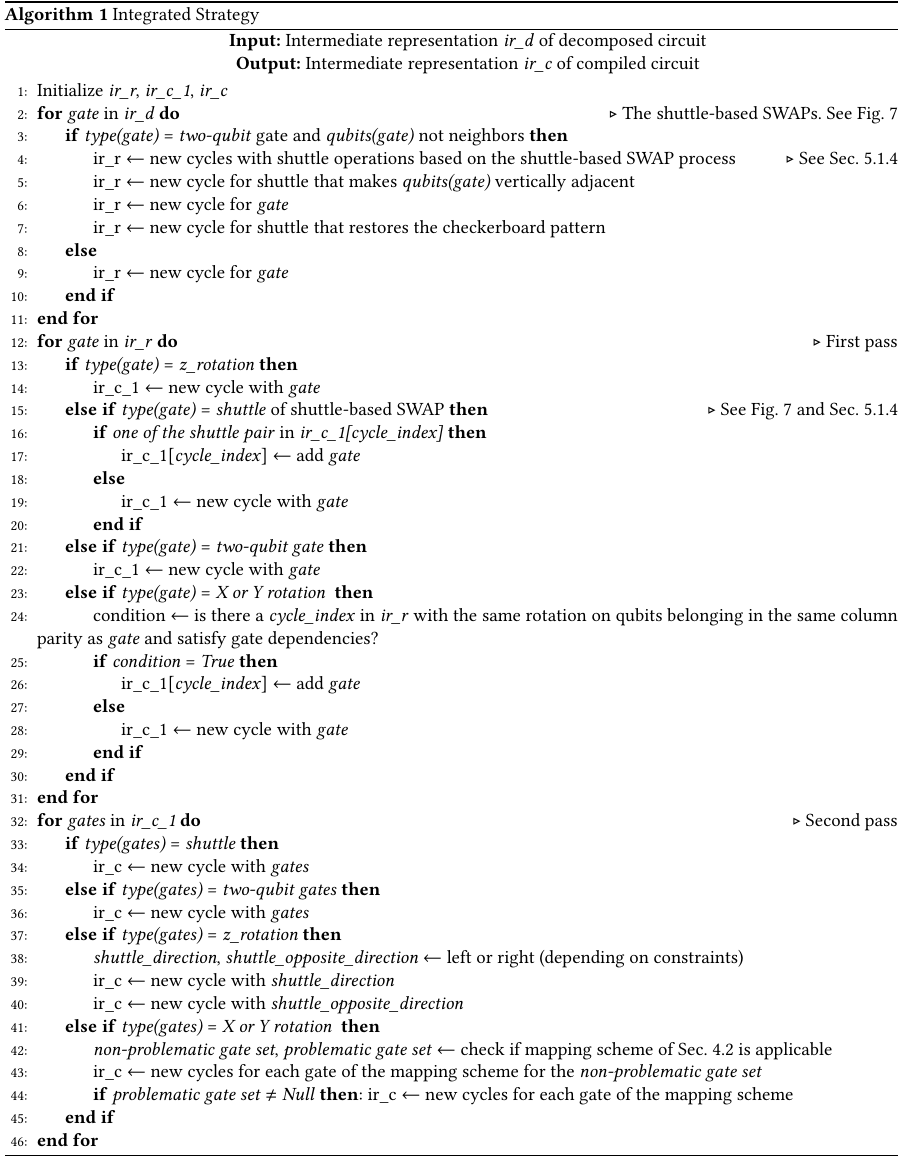}
\end{algorithm*}

As explained in Sec. \ref{Mapping challenges of a crossbar architecture}, both routing and scheduling techniques must avoid conflicts. To do that, a specific strategy needs to be conceptualized. There can be various strategies with various performance and compilation time trade-offs. The presented \textit{Integrated Strategy} tilts towards minimizing compilation time while having great prospects to be competitive against other strategies that focus on algorithm performance, as will be discussed in Sec. \ref{Discussion and future directions}.

To begin with, in the \textit{Integrated Strategy}, the checkerboard pattern qubit placement \cite{li2018crossbar}, also known as "idle-configuration" in \cite{helsen2018quantum}, should be maintained as much as possible. This provides at least two empty sites for every qubit to move towards to.

\begin{figure}
     \centering
     \begin{subfigure}[b]{0.5\textwidth}
         \centering
         \includegraphics[width=\textwidth]{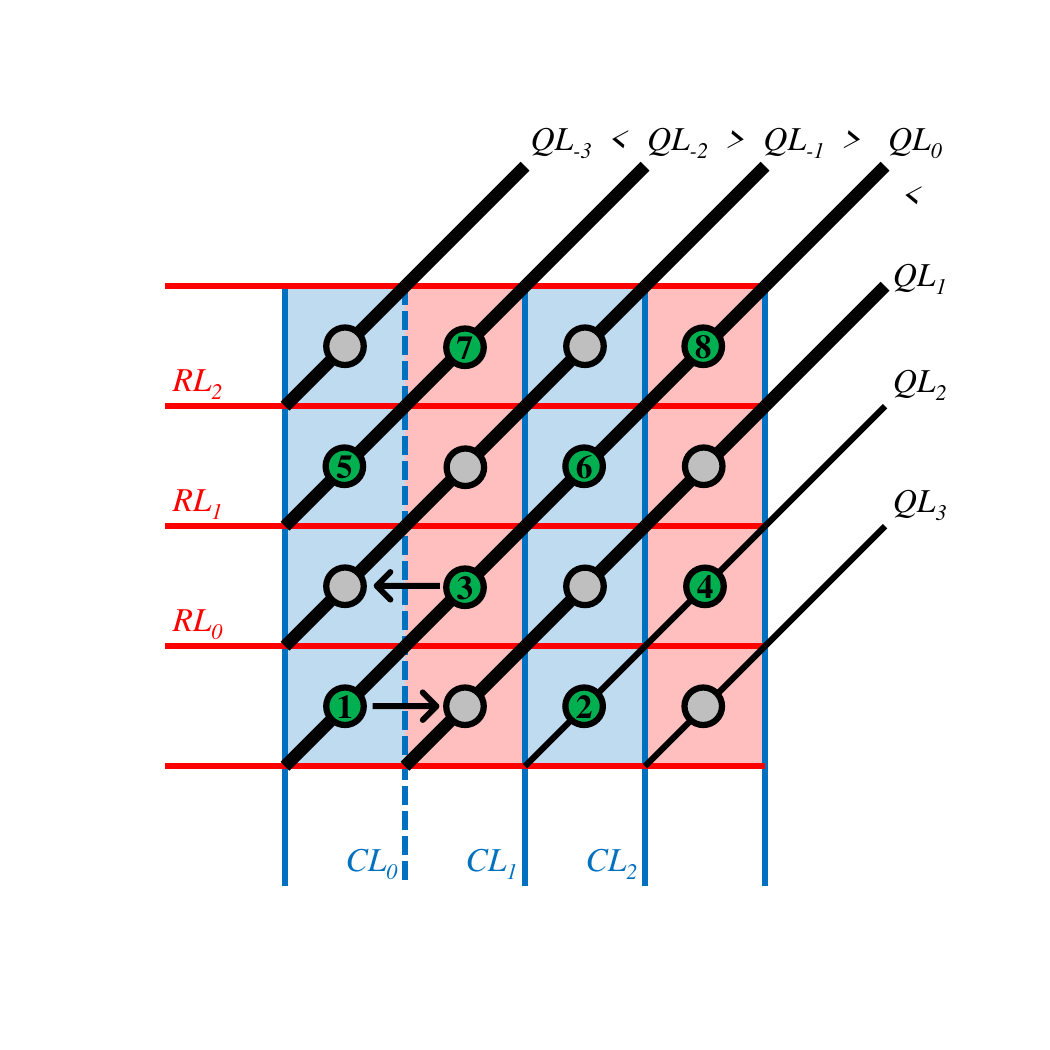}
         \caption{Horizontal shuttling}
         \label{fig:shuttle_based_SWAP_1}
     \end{subfigure}
    \hfill
     \begin{subfigure}[b]{0.5\textwidth}     
         \centering
         \includegraphics[width=\textwidth]{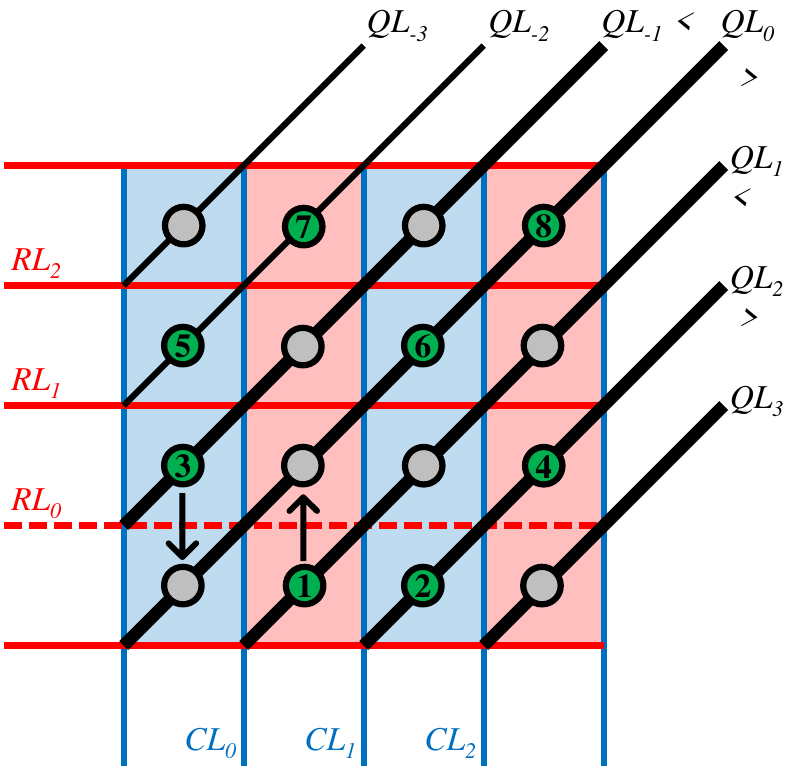}
         \caption{Vertical shuttling}
         \label{fig:shuttle_based_SWAP_2}
     \end{subfigure}
    \caption{Conflict-free shuttle-based SWAP for two-qubit gate routing: With this technique, two diagonally neighboring qubits exchange their position by consecutively performing two horizontal and two vertical shuttles. Each pair can be performed in parallel.}
    \label{fig:shuttle_based_SWAP}         
\end{figure}

When routing for two-qubit gates, we maintain the checker-board pattern throughout the circuit execution with a conflict-free shuttle-based SWAP technique \cite{morais2019mapping} as shown in Fig. \ref{fig:shuttle_based_SWAP}. Note that this movement of qubits results in a gate overhead of $4$ (i.e., $4$ shuttle operations), but a depth overhead of $2$, as these two shuttle pairs can always be executed in parallel. To bring one of the qubit operants to the appropriate position before the two-qubit interaction, multiple shuttle-based SWAPs might be performed. For that, we have implemented a shortest-path algorithm based on the Manhattan distance between the qubit operants. Once the two qubits are in the shortest position possible, the next step is a horizontal shuttle of one of them, either to the left or to the right, after which the target and control qubits are vertically adjacent, and the checkerboard pattern is temporarily broken. Proceeding the \(\sqrt{SWAP}\), a final shuttle returns the qubit to the previous position, and the checkerboard pattern gets restored. Note that the aforementioned process can be successfully executed only in that particular order, otherwise, there can be a routing conflict. Overall, routing for two-qubit gates requires at least one shuttle-based SWAP and exactly two horizontal shuttles.

So far, we have only talked about routing for bringing together qubits for performing two-qubit gates. However, qubit routing is also needed for shuttle-based Z rotations and might be needed for $X$ or $Y$ rotations, as discussed in Sec. \ref{Z-phase rotations} and Sec. \ref{Application of X or Y rotations on single qubits}, respectively. Mapping these two categories of gates, therefore, should always respect operational requirements and avoid conflicts. This also means that the "idle configuration" should be maintained when routing for these gates, as well. Thus, the second consideration of the \textit{Integrated Strategy} is the integration of single-qubit gate routing within the scheduling stage, hence the name "integrated", in order to prevent conflicts and optimize performance.

The \textit{Integrated Strategy} continues with two passes. In the first pass, the scheduler tries to parallelize $X$ or $Y$ gates in an ideal manner, \textcolor{black}{based on the gate dependencies \cite{lao2021timing},} (ignoring any potential conflicts) and Z gates individually. This is no different than other single-qubit gate scheduling processes proposed for other qubit architectures. However, it differs on the second pass which integrates the routing procedures for X, Y, and Z gates. The second pass iterates over each cycle produced by the first pass. For each cycle, there are two causes: \textcolor{black}{(a) if no conflicts are detected when scheduling the shuttle instructions of the mapping scheme described in Sec. \ref{Application of X or Y rotations on single qubits}, these instructions are inserted, each in a new cycle one after the other (b) if conflict(s) are detected, the subset of the problematic gate(s) is separated. Once the non-problematic gate subset is scheduled according to case (a), the problematic subset is recalled. This time it constitutes a conflict-free cycle and is scheduled similarly to case (a). The \textit{Integrated Strategy} is described in Algorithm 1 and its time complexity is calculated to be $O(n)$, with \textit{n} as the number of gates.}

\textcolor{black}{The key concept of this strategy is the ideal parallelization in the first pass, which is aimed to relieve the increased complexity of concurrently avoiding conflicts and optimizing. Then, the second pass tries to satisfy the scheduling of the first pass in the least cycles possible while completing the mapping steps of all gates (e.g., applying the mapping scheme for X or Y gates or adding shuttles for Z rotations, etc.). Overall, this first implementation of our \textit{Integrated Strategy} does not parallelize gates of different types in the same cycle, and thus each cycle is dedicated to one instruction type. Additionally, it leaves room for improvement while maintaining its $O(n)$ time complexity, as discussed in Sec. \ref{Integrated strategy improvements}. Fortunately, the strategy described above and suggested extensions in Sec. \ref{Integrated strategy improvements} can be adapted to a real setup. As explained in Sec. \ref{Mapping challenges of a crossbar architecture}, a fabricated crossbar device will most likely have material imperfections, thus requiring pulse calibration per site. As pointed out by \cite{helsen2018quantum,kharkov2022arline}, pulsing control lines prematurely to account for material variations could cause an unwanted interaction. Since, however, the \textit{Integrated Strategy} (or an extension thereof) exclusively schedules gates of the same type in each cycle, fine-tuning pulses within is possible before moving to the next cycle.}

\subsection{Performance metrics}

We will now introduce the metrics used in this work to evaluate the performance of \textit{SpinQ} when mapping different algorithms on the crossbar architecture.

\subsubsection{\textbf{Gate overhead}}
One commonly used metric to evaluate the performance of a mapper and its underlying architecture is gate overhead. We calculate it as the percentage relation of additional gates inserted by the mapper to the number of gates after decomposition. We do not count decomposition gate overhead as it is always proportional to the number of gates. Getting a clear view of the various sources of gate overhead will help to form useful insights. Therefore, the main sources of gate overhead are the following:

\begin{itemize}
\item $4$ additional shuttle instructions per shuttle-based SWAP for two-qubit gates
\item At least $3$ additional instructions for each $X$ or $Y$ rotation gate due to the semi-global rotation scheme (Sec. \ref{Application of X or Y rotations on single qubits})
\item $2$ additional shuttle instructions for each two-qubit gate 
\item $1$ additional shuttle operation for each Z rotation gate
\end{itemize}

Note that, unlike superconducting architectures where gate overhead results from routing instructions (i.e., SWAP gates) for performing two-qubit gates, in the crossbar, it can be caused by single-qubit gates as well.

\subsubsection{\textbf{Depth overhead}} \label{Depth overhead}
Another commonly used metric to evaluate the performance of a mapper and its underlying architecture is the depth overhead of a circuit. The depth of a circuit is equal to the minimum number of time steps of a circuit when executing gates in parallel \cite{sinha2022qubit,lao2021timing,bandic2020structured,zulehner2018efficient,herbert2018using}. Note that the initial circuit depth is calculated after scheduling the circuit only by its gate dependencies, meaning without any architectural constraints. We calculate depth overhead as the percentage relation of additional depth produced by the mapper to the circuit depth after decomposition. The main sources of depth overhead are: 

\begin{itemize}
\item At least $3$ additional cycles for each $X$ or $Y$ rotation gate due to the semi-global rotation scheme (Sec. \ref{Application of X or Y rotations on single qubits})
\item $2$ additional cycles per shuttle-based SWAP for two-qubit gates
\item $2$ additional cycles for each two-qubit gate
\item $1$ additional cycle for each Z rotation gate
\end{itemize}

\subsubsection{\textbf{Estimated Success Probability}}
A key metric to assess the performance not only of the compiler but in general of a quantum computing system is the algorithm's success rate.  From an experimental point of view, the algorithm success rate is calculated by executing the algorithm several times on a given (real) quantum processor and creating the distribution of successful executions, based on the expected measurement. \textcolor{black}{ An alternative way to calculate the success rate without the need for a real quantum processor is by classical simulation. Recently, hybrid Schrodinger-Feynman simulations have been used efficiently, but only for shallow circuits \cite{pan2022simulation,burgholzer2021hybrid,markov2018quantum}. Another method uses tensor networks and has been shown as a more scalable technique for IBM’s Eagle or Google's Sycamore chips, but suffers from exponential time complexity on more connected architectures or circuits larger in both depth and width  \cite{pan2022simulation,tindall2023efficient}. }

\textcolor{black}{However, there is a need for a more efficient method able to approximate the success rate of much larger circuits. One of the most commonly used methods is considering the final compiled circuit and particular architectural configurations given as input at the beginning of compilation \cite{quetschlich2022predicting, Diogo}. The estimated success probability (ESP) of an algorithm can be calculated as:}

\begin{equation}
ESP = \prod_{i}\prod_{j}{gate\char`_fidelity_{i,j}}
\label{eq:ESP_0}
\end{equation}

where $i$ represents the $i$th time step and $j$ the $j$th gate in the $i$th time step.

\textcolor{black}{This method is less complex in both time and space, but not as accurate, compared to classical simulations. From equation \ref{eq:ESP_0} it is evident that the time complexity of ESP only increases linearly with the number of gates in a circuit while space complexity remains constant, contrary to the aforementioned methods.} To expand it, we have considered a per-type and per-location variability of gate fidelities based on a normal distribution. This implies that, for instance, a two-qubit gate (e.g., \(\sqrt{SWAP}\)) will have lower fidelity than a single-qubit gate and that the actual fidelity will depend on the exact location in the topology. These expansions constitute a more realistic, i.e., closer to a real device, estimation of circuit success probability:
\begin{equation}
    ESP = \prod_{i}\prod_{j}{gate\char`_fidelity_{i,j}^{x,y}}
    \label{eq:ESP_1}
\end{equation}

where $i$ represents the $i$th time step, $j$ the $j$th gate in the $i$th time step and and $x,y$ are the physical qubit(s) coordinates.

\subsubsection{\textbf{Compilation time}
}In this work, we are not only interested in building mapping techniques themselves but also in their scalability potential. This necessitates that our proposed \textit{SpinQ} strategy should remain efficient for a variety of quantum circuit parameters (e.g., number of qubits or percentage of two-qubit gates). By measuring the compilation time for mapping quantum circuits, we get a reference of the scalability of our implementations.

\subsection{Verification}

A verification tool is important to this work due to the lack of a working device for real-system testing. \textcolor{black}{It is used on demand in the initial stage of development to debug and verify current or future mapping approaches. The tool searches for mismatches between the qubits' position history stored during the compilation and all shuttling sequences. This ensures that all routing instructions added in the final compiled circuit will shuttle the right qubits in the correct places without conflicts, and vice versa. This is critical for an architecture such as this one where both the routing and scheduling can produce conflicts. It also checks for operational constraint violations and potential conflicts caused by those. Finally, and after the previous checks, a state vector simulation takes place between the main stages of the compiler with the use of Qiskit Aer library \cite{Aer}. Specifically, it compares the probability distributions produced by the \textit{qasm\_simulator} backend between the initial circuit, the decomposed, the routed for two-qubits gates, and the one processed by the \textit{Integrated Strategy}. This ensures that the mapping techniques and the compilation strategies used do not change the outcome of the algorithm. However, it should be noted that in non-application-based algorithms (e.g., randomly generated) their state distribution probability can be anything and will suffer a change just from the decomposition stage compared to application-based algorithms. For this reason, this last verification stage can not be used for all benchmarks. Additionally, this verification can not be used for more than $30$ qubits due to exceeded memory requirements.}

\section{Experimental Methodology} \label{Experimental Methodology}

\subsection{Benchmarks}
We have generated $3,630$ random uniform algorithms \cite{sivarajah2020t} containing X, Y, Z and $\sqrt{SWAP}$ gates (all native to the crossbar architecture) to be used as benchmarks. With this set, we can vary on demand the number of gates, number of qubits, and percentage of two-qubit gates. For example, a random uniform benchmark with $50\%$ of two-qubit gates relative to single-qubit gates will have $33.33\%$ of $X$ or $Y$ gates, $33.33\%$ of Z gates, and $33.33\%$ of two-qubit gates. Generating synthetic circuits provides a well-controlled benchmark collection from which we can better understand results and form insights. Moreover, we use real benchmarks from the RevLib library in a [5 - 1400] gate range \cite{wille2008revlib}. Quantum circuits from this library are often used in related quantum circuit compilation works \cite{lao20222qan,zulehner2018efficient,murali2019noise} and it consists of quantum algorithms with parameters ranging from $3$ to $16$ qubits, $18.75\%$ to $100\%$ of two-qubit gates and $5$ to $512,064$ gates. Finally, we also consider quantum circuits from the Qlib library \cite{lin2014qlib} which contains real quantum algorithms in increasing sizes.

\subsection{Benchmarks characterization}
When it comes to performance evaluation, it is important to not only consider properties of the architecture but also the characteristics of quantum circuits. The simplest and most commonly \cite{bandic2022full} used parameters of quantum circuits are number of qubits, number of gates, and absolute or relative (i.e., percentage) number of two-qubit gates. However, only these three characteristics can be misleading for two reasons. Firstly, two benchmarks, for instance, could have the same parameter values but heavily differ in the circuit's structure \cite{bandic2022full}. When one of them has all pairs of qubits interact with each other, it will require more routing than the other, which might have the same number of interactions, but with only one pair of qubits interacting. The structure of a quantum circuit is derived from its qubit interaction graph (QIG) which represents the number and distribution of interactions (i.e., two-qubit gates) between virtual qubits. Several internal circuit parameters can be extracted from the QIG that better distill its properties \cite{bandic2022full}. Having said that, we analyze QIGs mainly visually, as this is still an active field of research \cite{bandic2022full}. \textcolor{black}{We support these conclusions by extracting the average \textit{degree} \cite{bandic2022full} or \textit{program communication} \cite{tomesh2022supermarq}  of the QIG, which represents the average number of edges that are incident to (i.e., connected to) a node. In simple terms, it expresses the level of "connectedness" of a graph. We can thus make concrete conclusions and form insights from such a QIG assessment.} The second reason is that initial gates can be decomposed to natively supported instructions for the underlying architecture. This means that the number of gates and ratios (percentages) between each gate type can differ from the initial set to the decomposed set, meaning that evaluations can become more accurate when accounting for the decomposed set. 

\subsection{Experimental Setup}
We run \textit{SpinQ} on a laptop with an Intel(R) Core(TM) i7-3610QM CPU @ ~3.20GHz and 16GB DDR3 memory. \textit{SpinQ} is written in Python 3.9.6 version.

\section{Evaluation and analysis} \label{Evaluation and analysis}

In this Section, we present an in-depth performance analysis of \textit{SpinQ} when mapping a broad range of quantum algorithms on the crossbar architecture. We then form architectural and mapping insights for each performance metric. More specifically, gate overhead and corresponding insights are presented in Sec. \ref{Gate Overhead} and \ref{Insights from gate overhead analysis}, depth overhead in Sec. \ref{Depth Overhead} and \ref{Insights from depth overhead analysis}, and ESP in Sec. \ref{Estimated Success Probability} and \ref{Insights from Estimated Success Probability analysis}. Finally, we show results regarding the compilation time of \textit{SpinQ} in Sec. \ref{Compilation time} to asses its scalability capability.

\subsection{Gate Overhead} \label{Gate Overhead}

\begin{figure}
     \begin{subfigure}[b]{0.5\textwidth}
         \centering
         \includegraphics[width=\textwidth]{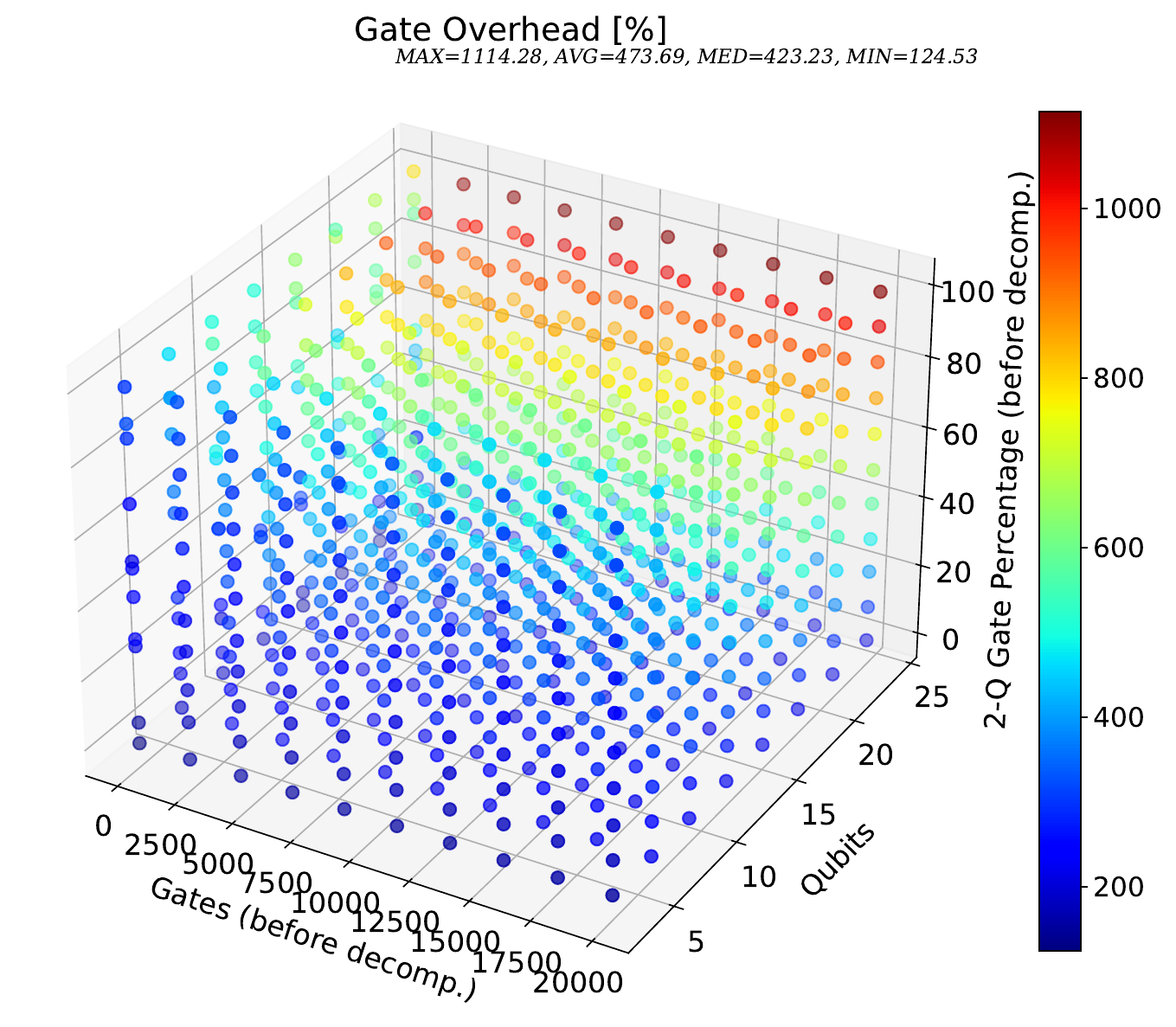}
         \caption{}
         \label{fig:gaterandom_small}
     \end{subfigure}
     \begin{subfigure}[b]{0.5\textwidth}
         \centering
         \includegraphics[width=\textwidth]{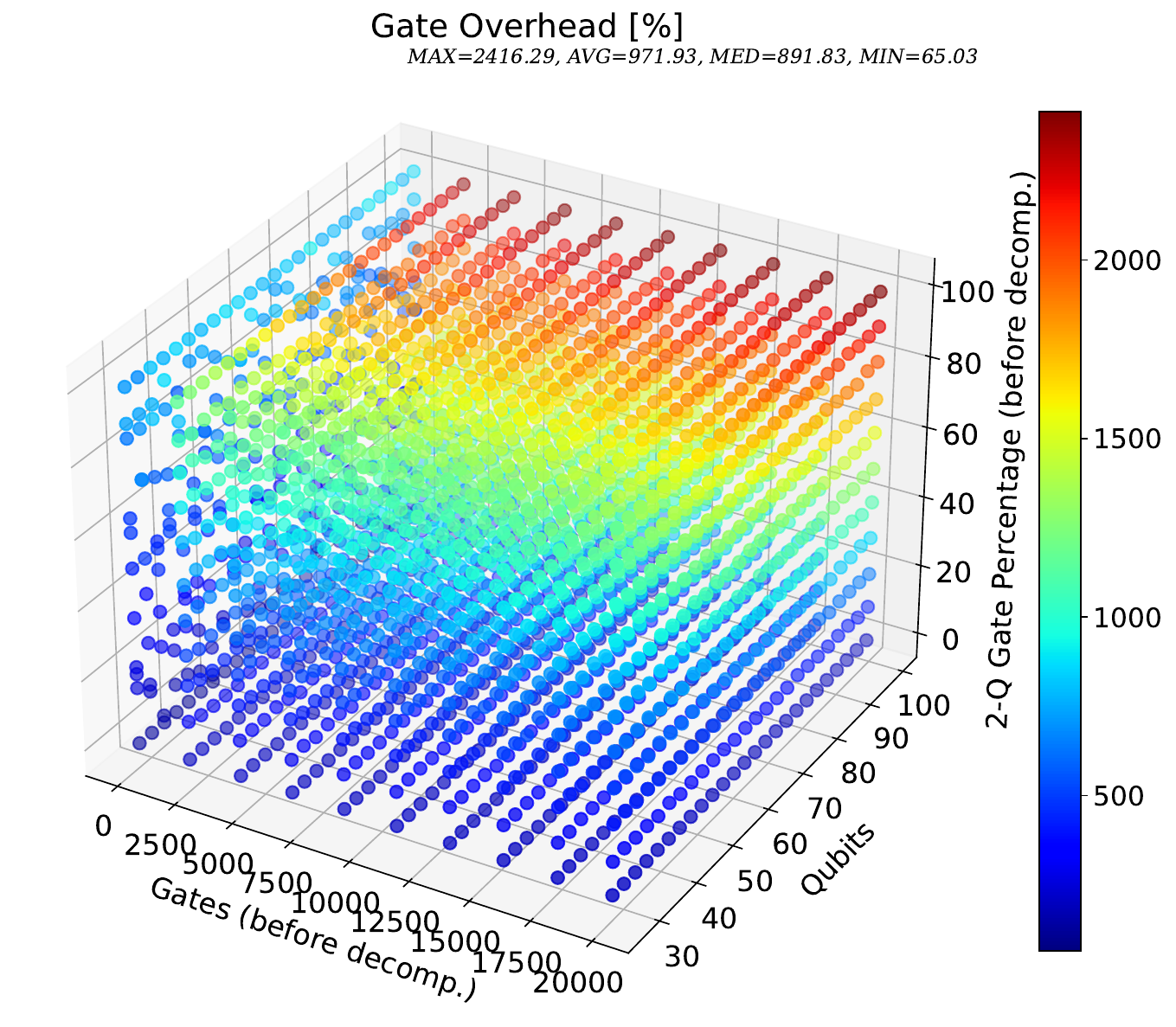}
         \caption{}
         \label{fig:gaterandom_medium}
     \end{subfigure}  
\caption{Resulting gate overhead when $3,630$ random uniform quantum algorithms are mapped onto the crossbar architecture. The three axes correspond to benchmark characteristics, namely, the number of gates [50 - 20,000], number of qubits [3 - 99] (split into two subfigures), and two-qubit gate percentage [0 – 100].}
\label{fig:gaterandom}
\end{figure}

\begin{figure}[htpb]
     \begin{subfigure}[b]{0.5\textwidth}
         \centering
         \includegraphics[width=\textwidth]{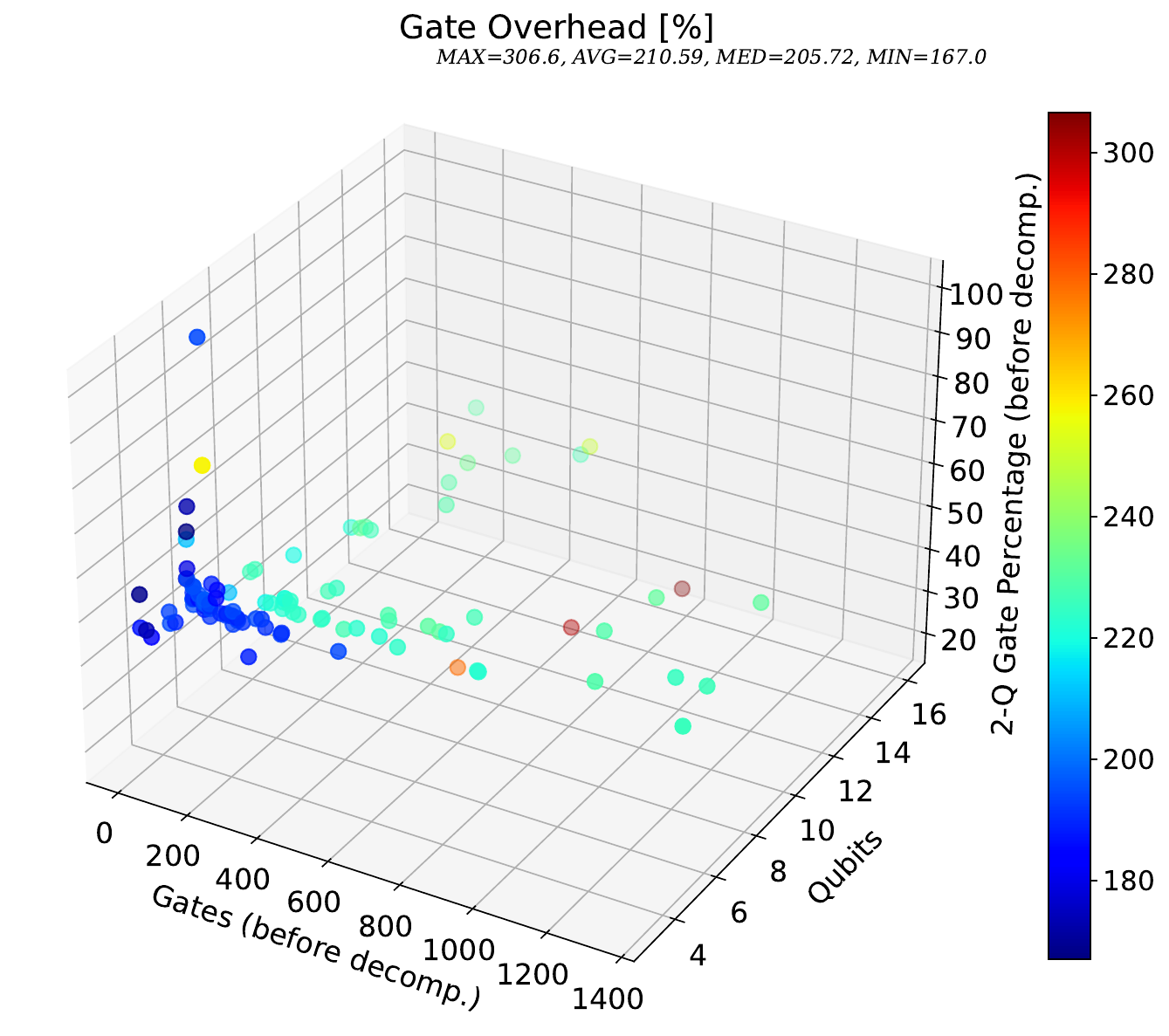}
         \caption{}
         \label{fig:gaterevlib}
     \end{subfigure}
     \begin{subfigure}[b]{00.5\textwidth}
         \centering
         \includegraphics[width=\textwidth]{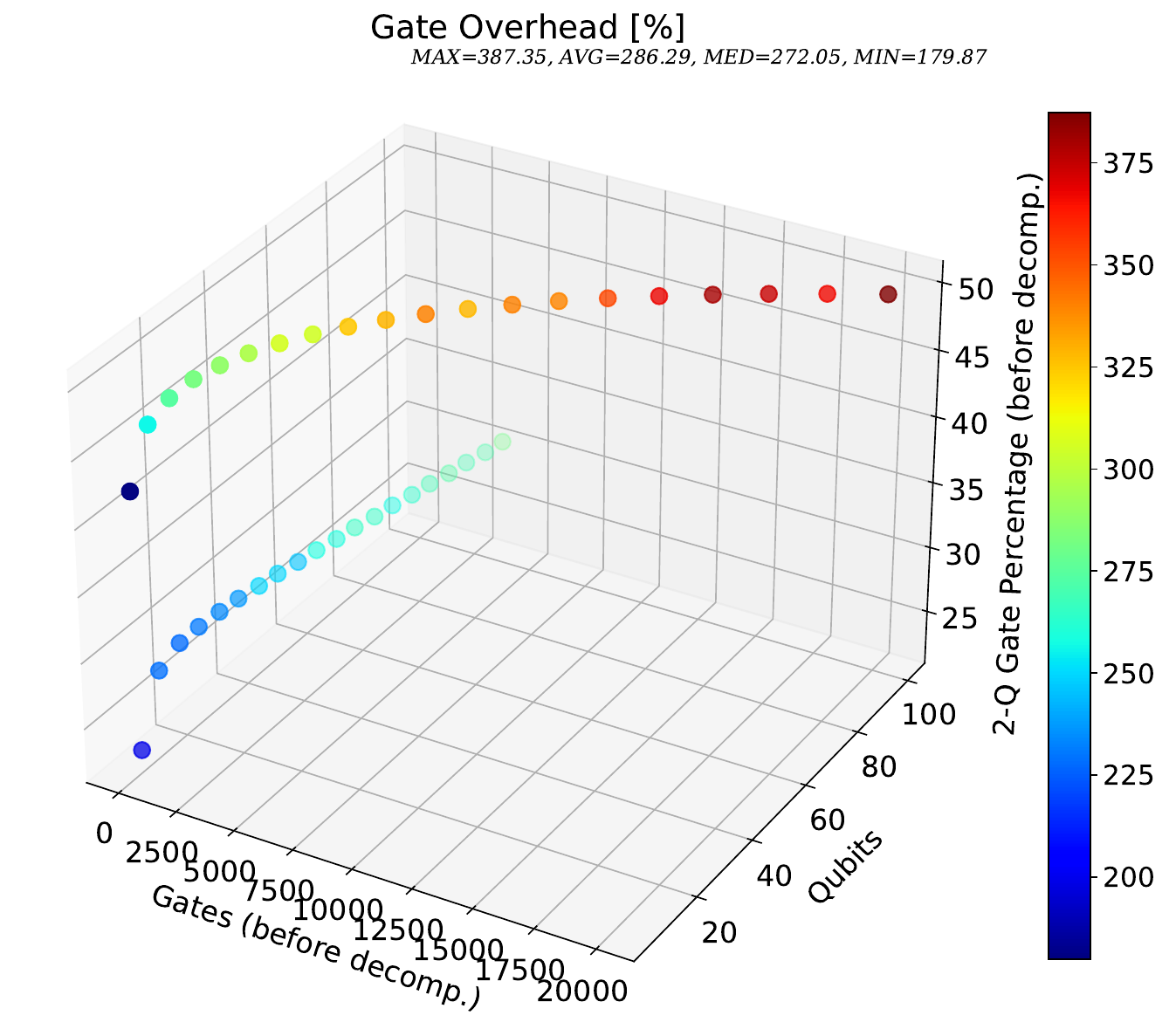}
         \caption{}
         \label{fig:gate_qft_grover}
     \end{subfigure}  
  \caption{\textcolor{black}{(a) Resulting gate overhead when mapping quantum algorithms from the RevLib library onto the crossbar architecture. The three axes correspond to benchmark characteristics, namely, number of gates [5 - 1400], number of qubits [3 - 16], and two-qubit gate percentage [18.75 - 100]. RevLib algorithms consist of reversible \cite{bandic2022interaction} quantum algorithms including, but not limited to, arithmetic and encoding functions \cite{WGT+:2008}. (b) Resulting gate overhead when mapping Grover's (bottom line of data points) and QFT (top line of data points) quantum algorithms onto the crossbar architecture. The three axes correspond to benchmark characteristics, namely, number of gates [52 - 20050], number of qubits [5 - 100] and two-qubit gate percentage [22.86 - 49.63].}}
  \label{fig:gaterevlib}  
\end{figure}

\begin{figure}[htpb]
    \centering
    \includegraphics[width=0.5\textwidth]{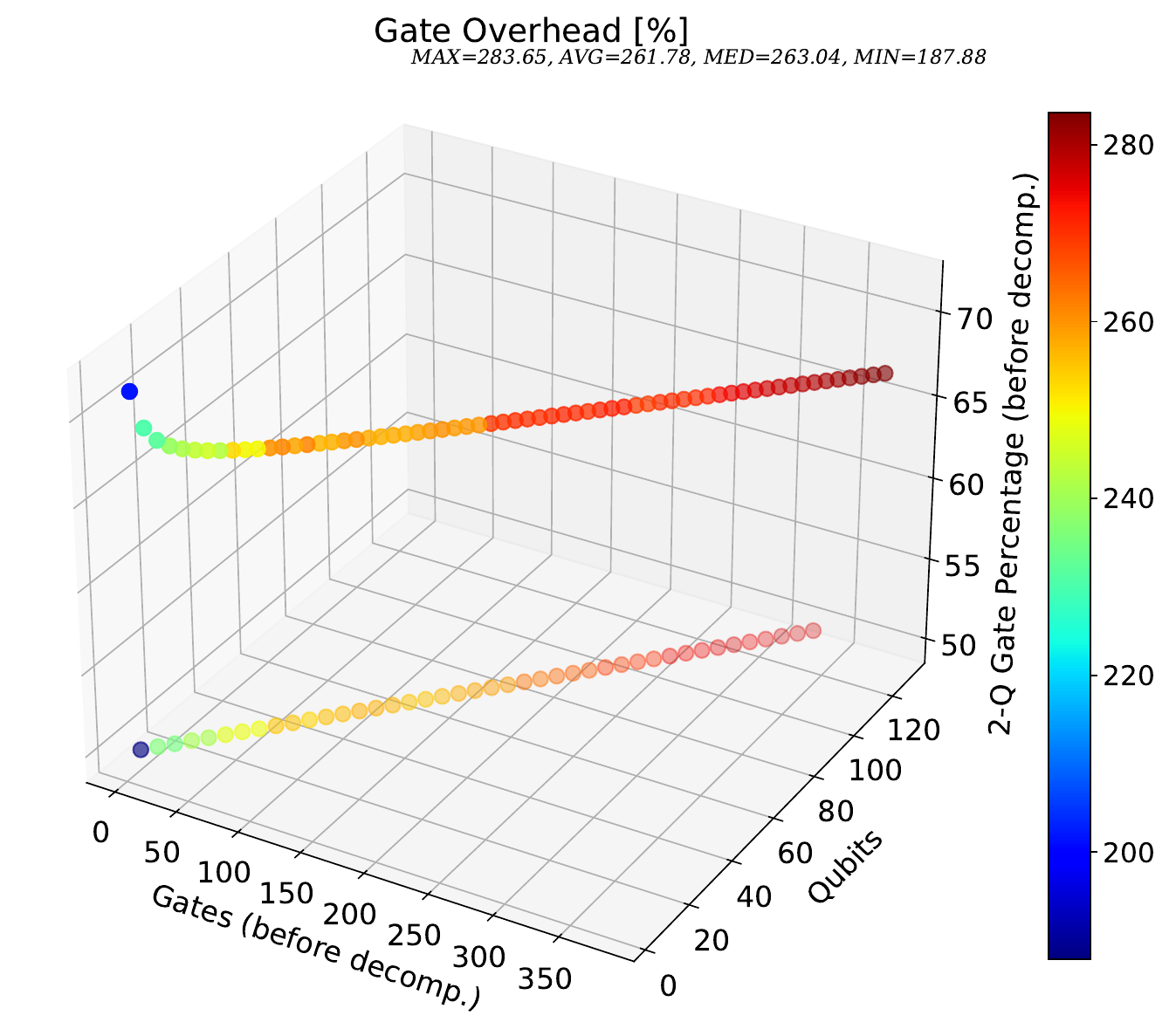}
    \caption{Resulting gate overhead when mapping the Cuccaro Adder (top line of data points) and the Vbe Adder (bottom) quantum algorithms from the Qlib library onto the crossbar architecture. The three axes correspond to benchmark characteristics, namely, number of gates [4 - 385], number of qubits [4 - 130] and two-qubit gate percentage [50 - 71.43].}
    \label{fig:gatevbecucCOMP}
    
\end{figure}

\begin{figure*}
     \includegraphics[width=\textwidth]{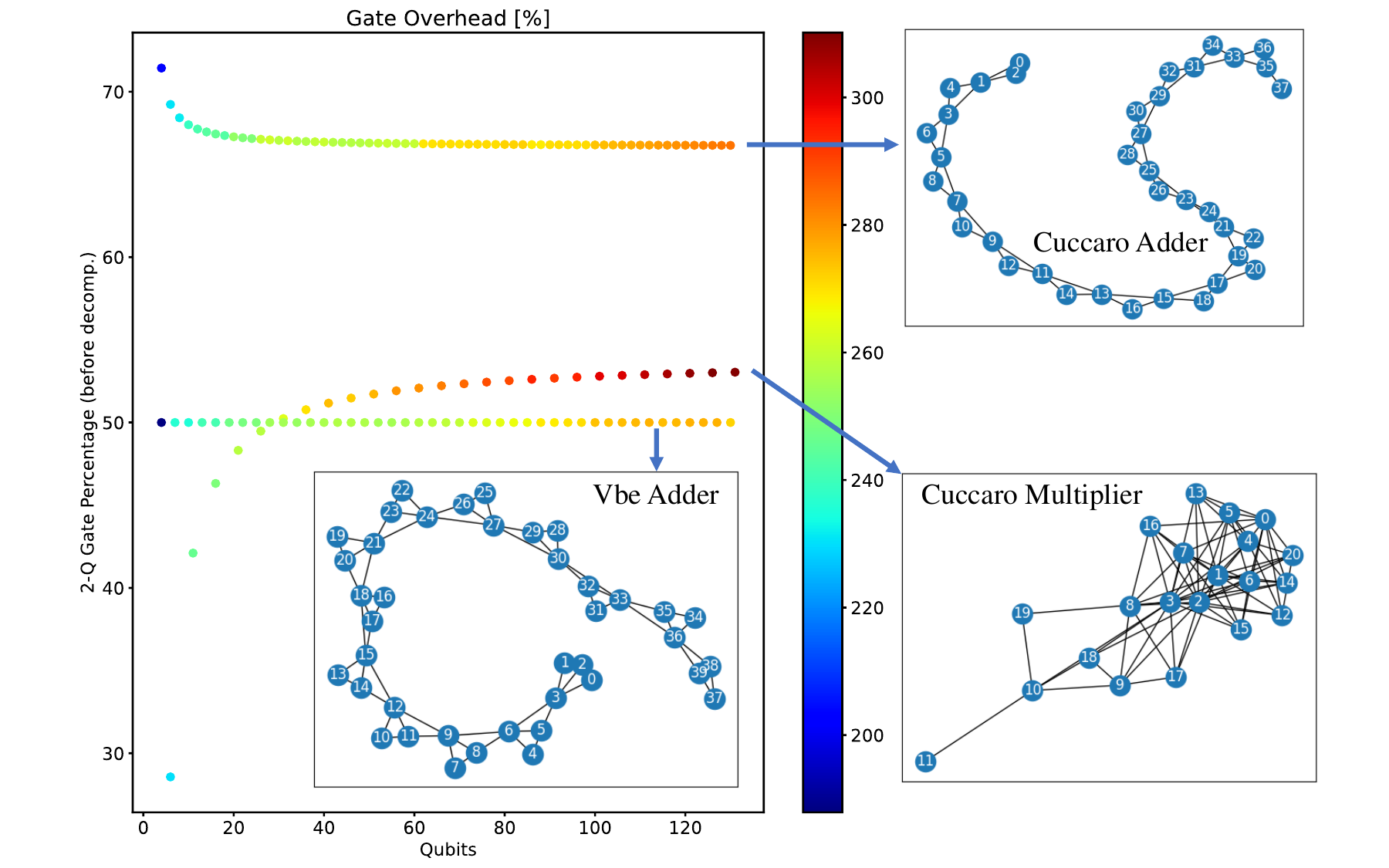}
    \caption{Resulitng gate overhead when the Vbe Adder, Cuccaro Adder, and Cuccaro Multiplier from the Qlib library are mapped onto the crossbar architecture alongside their Quantum Interaction Graphs (QIG) consisting of 40, 38, and 21 qubits, respectively. The y-axis represents the two-qubit gate percentage and the x-axis the number of qubits. We see gate overhead to be influenced not only by the number of qubits and two-qubit gate percentage but also by the qubit interaction distribution.}
  \label{fig:gate_cuc_vbe_multiplier_interaction}
\end{figure*}

To start with, we analyze the gate overhead trend in a wide range of quantum algorithms. In Fig. \ref{fig:gaterandom} we have mapped random uniform circuits on the crossbar architecture. Focusing on Fig. \ref{fig:gaterandom_small}, which reaches up to $25$ qubits, we observe that as we go from low to high number of qubits and from low to high percentage of two-qubit gates, the gate overhead increases (from blue to red color). More precisely, higher qubit counts imply larger crossbar topologies, thus potentially longer routing distances, i.e., more shuttle-based SWAPs. Furthermore, higher percentages of two-qubit gates potentially lead to more routing of qubits. These observations verify that the main source of gate overhead is indeed the routing of qubits for two-qubit gates (see Sec. \ref{Compiler Stages}). We also notice that the number of gates has a small but noticeable influence on the gate overhead. To further observe the trend when increasing the number of qubits, we changed the range of qubits from [3 – 25] to [25 – 99] in Fig. \ref{fig:gaterandom_medium}. We see once more that the gate overhead increases as we go from low to high number of qubits and percentage of two-qubit gates. As expected, the gate overhead, shown on the color bars, of the [25 – 99] qubit range is on average $102.49\%$ higher than that of the [3 – 25] qubit range because of the increased routing distances.

So far, the above random algorithms were generated to have control of different circuit parameters (i.e., number of qubits and gates and two-qubit gate percentage) in a way to broadly cover the parameter space and up to certain boundaries. However, they might not be representative of real algorithms from a circuit structure point of view (e.g., how two-qubit gates are distributed among qubits or the degree of operation parallelism). Therefore, we then mapped real algorithms from the RevLib \textcolor{black}{\cite{WGT+:2008}} and Qlib \textcolor{black}{\cite{lin2014qlib}} libraries, Grover and QFT resulting in the gate overhead shown in Fig. \ref{fig:gaterevlib}, Fig. \ref{fig:gatevbecucCOMP}, and Fig. \ref{fig:gate_cuc_vbe_multiplier_interaction}. In Fig. \ref{fig:gaterevlib} (a) we can observe that benchmarks “cluster” together in similar colors, namely shades of blue, green, yellow, and red. This implies that similar benchmarks, meaning with similar parameters and structure, have similar gate overhead. Note that whereas random uniform algorithms have the same circuit structure because of the way they are generated, RevLib algorithms present different structural parameters not only compared to the randomly generated circuits but also between them. For this reason, correlations such as the higher the number of qubits and two-qubit gate percentage gets, the higher the gate overhead, are not as evident as for the random circuits. \textcolor{black}{Similarly, in Fig. \ref{fig:gaterevlib} (b) we have executed Grover's and QFT algorithms. With these simulations, we also want to perform a scalability analysis of algorithms, which is not possible with RevLib circuits. From a first observation, it seems that QFT (top dots) produces higher gate overhead due to the higher two-qubit gate percentage compared to Grover's algorithm. However, once again, this can not be conclusive as they scale in different rates of benchmark characteristics.} 

To further analyze how structural circuit parameters impact the gate overhead, we mapped algorithms with similar rates of number of gates, qubits, percentage of two-qubit gates and QIGs. First, note in Fig. \ref{fig:gatevbecucCOMP} that the Cuccaro Adder (top line in Fig. \ref{fig:gatevbecucCOMP}) has a small drop in the percentage of two-qubit gates that goes from $71.43\%$ to $66.75\%$ when increasing in size (number of qubits) whereas the Vbe Adder (bottom line) maintains a lower percentage of 50\% for the same qubit increase. One can immediately observe that the Cuccaro Adder shows a higher gate overhead up to $284\%$ due to the higher two-qubit gate percentage compared to the $271\%$ of Vbe Adder, matching the conclusions made for Fig. \ref{fig:gaterandom}. However, as we emphasized above, in the case of real algorithms comparisons can only be properly made when looking not only at their circuit parameters but also at their more structural ones such as the QIG.

For this reason, in Fig. \ref{fig:gate_cuc_vbe_multiplier_interaction} we show the derived QIGs from Vbe Adder's 40-qubit circuit, Cuccaro Adder’s 38-qubit circuit and Cuccaro Multiplier's 21-qubit circuit alongside their gate overhead in relation to the number of qubits and two-qubit gate percentage. In these QIGs, nodes correspond to qubits and edges to qubit interactions, i.e.,  two-qubit gates. The particular QIGs size selection was made to easily show their structure. We immediately observe similarities in the QIGs of the two Adders as the distribution of interactions is almost identical. More specifically, we see $2$ to $3$ interactions per qubit on average, with others close to their logical qubit number. \textcolor{black}{Such a visual observation can be also quantified with the average QIG \textit{degree}, which is calculated to be $3$ for both. It is not surprising, therefore, that the higher gate overhead of Cuccaro Adder is indeed due to the higher percentage of two-qubit gates, compared to Vbe Adder. }

However, note that the Cuccaro Multiplier has the highest gate overhead of all three ($309\%$) despite having a lower two-qubit gate percentage than the Cuccaro Adder. Looking at its much more connected QIG implies a denser qubit interaction distribution, compared to the others. \textcolor{black}{Its average \textit{degree} is determined to be $8$; higher than that of the two other algorithms.} Because of this, more routing is needed to connect nearly all qubits across the topology. 

\subsection{Insights from gate overhead analysis} \label{Insights from gate overhead analysis}

Accounting for the routing constraints, as discussed in Sec. \ref{Mapping challenges of a crossbar architecture}, mapping on the crossbar architecture is not a trivial task. In fact, we have emphasized the importance of conceptualizing and developing new routing techniques that specifically can address the unique mapping challenges of spin-qubit architectures. More specifically, with the adoption of the checkerboard pattern combined with the shuttle-based SWAPs, we can provide a scalable solution of qubit routing for two-qubit gates. Additionally, the complexity only scales with the number of two-qubit gates, therefore being a viable solution for large-scale implementation. However, this technique makes two-qubit gate routing the highest source of gate overhead and it can dramatically increase it with higher qubit counts and a higher percentage of two-qubit gates (see Fig. \ref{fig:gaterandom} and \ref{fig:gatevbecucCOMP}). Moreover, in Fig. \ref{fig:gate_cuc_vbe_multiplier_interaction} we saw that gate overhead can also be increased by a more connected QIG even though other circuit parameter values are comparatively lower. This shows the importance of basing circuit performance evaluation not only on simple circuit parameters but also on other `hidden' structural characteristics such as the qubit interaction distribution.

Having said that, the second biggest source of gate overhead originates from $X$ or $Y$ qubit rotations. This is due to the unprecedented semi-global rotation scheme. \textcolor{black}{As mentioned in Sec. \ref{Application of X or Y rotations on single qubits}, this first time that single-qubit gate mapping requires additional routing instructions (i.e., produce gate overhead) compared to other qubit architectures. In comparison, neutral-atom architecture \cite{graham2021demonstration,xia2015randomized,sheng2018high,levine2019parallel,patel2022geyser} demonstrated small algorithms execution by constructing local $R_\phi$ rotations at an arbitrary angle or axis by synthesizing a local $R_z$ in between two global $R_{xy}$. This is a similar concept to the semi-global rotation mapping scheme of the crossbar architecture but it differs in two key aspects. Firstly, one has the ability of global rotation whereas the other is only semi-global. This could be an advantage or a disadvantage in terms of performance depending on the algorithm at hand and on the mapping techniques used. Secondly and more importantly, the crossbar architecture demands a more constrained scheme that necessitates routing operations. These operations must be executed meticulously to address the distinct architectural restrictions, such as ensuring the availability of empty sites for qubit movement, adhering to the shuttling signal constraints, and diligently sidestepping any potential conflicts. Therefore, this feature is exclusive and raises unique mapping challenges for spin-qubit architectures and calls for careful considerations during the compilation process, as explained in Sec. \ref{Mapping challenges of a crossbar architecture}.} 

The previous two facts inspire novel mapping techniques for the crossbar architecture and potentially for other spin-qubit architectures with similar characteristics that can increase performance, namely:

\begin{enumerate}
    \item Developing a routing solution dedicated to accounting for potential conflicts and constraints can reduce the gate overhead resulting from the shuttle-based SWAPs. Such a generalized routing algorithm could also include SWAP interactions (two consecutive \(\sqrt{SWAP}\)s) and $CPHASE$ interactions. For instance, there can be scenarios that choosing a more noisy two-qubit interaction, for the purpose of avoiding an upcoming conflict, could result in higher ESP. Additionally, such a heuristic algorithm can allow multiple control or target qubits (\cite{lao2021timing}) to be shuttled around the topology enabling for parallelization of many two-qubit gates while avoiding high error variabilities in the topology \cite{tannu2019not}. However, such a solution must be implemented with the complexity in mind such that it will not make it unviable on large scale. 
    
    \item A more efficient routing algorithm for single-qubit gates can significantly reduce the gate overhead, such that a specific rotation scheme to rotate targeted qubits is used less often. Such an algorithm can route qubits to the appropriate odd or even columns before the execution of single-qubit gates eliminating the need to apply any scheme afterward, such as the one in Sec. \ref{Application of X or Y rotations on single qubits}.
    
    \item Combining the previous two points, there can be a unified algorithm implementing both. In such an algorithm, upcoming routing for single-qubit gates is accounted for when routing for two-qubit gates, and vice versa.
    
    \item Finally, an initial placement algorithm can take into account not only two-qubit gates but single-qubit gates as well. Since the positions of qubits influence the gate overhead resulting from single-qubit gate mapping (due to the semi-global rotation scheme), an extension of an initial placement algorithm accounting for single-qubit gates can further reduce the gate overhead.
\end{enumerate}
 
Last but not least, we have emphasized that to concretely evaluate results, there has to be sufficient characterization of benchmarks, especially when evaluating novel architectures and mapping techniques. In our analysis, we did not rely only on simple benchmark parameters, such as the percentage of two-qubit gates, but also on the internal structure of benchmarks using the Quantum Interaction Graph (QIG).

\subsection{Depth Overhead} \label{Depth Overhead}

\begin{figure}
     \centering
     \begin{subfigure}[b]{0.5\textwidth}
         \centering
         \includegraphics[width=\textwidth]{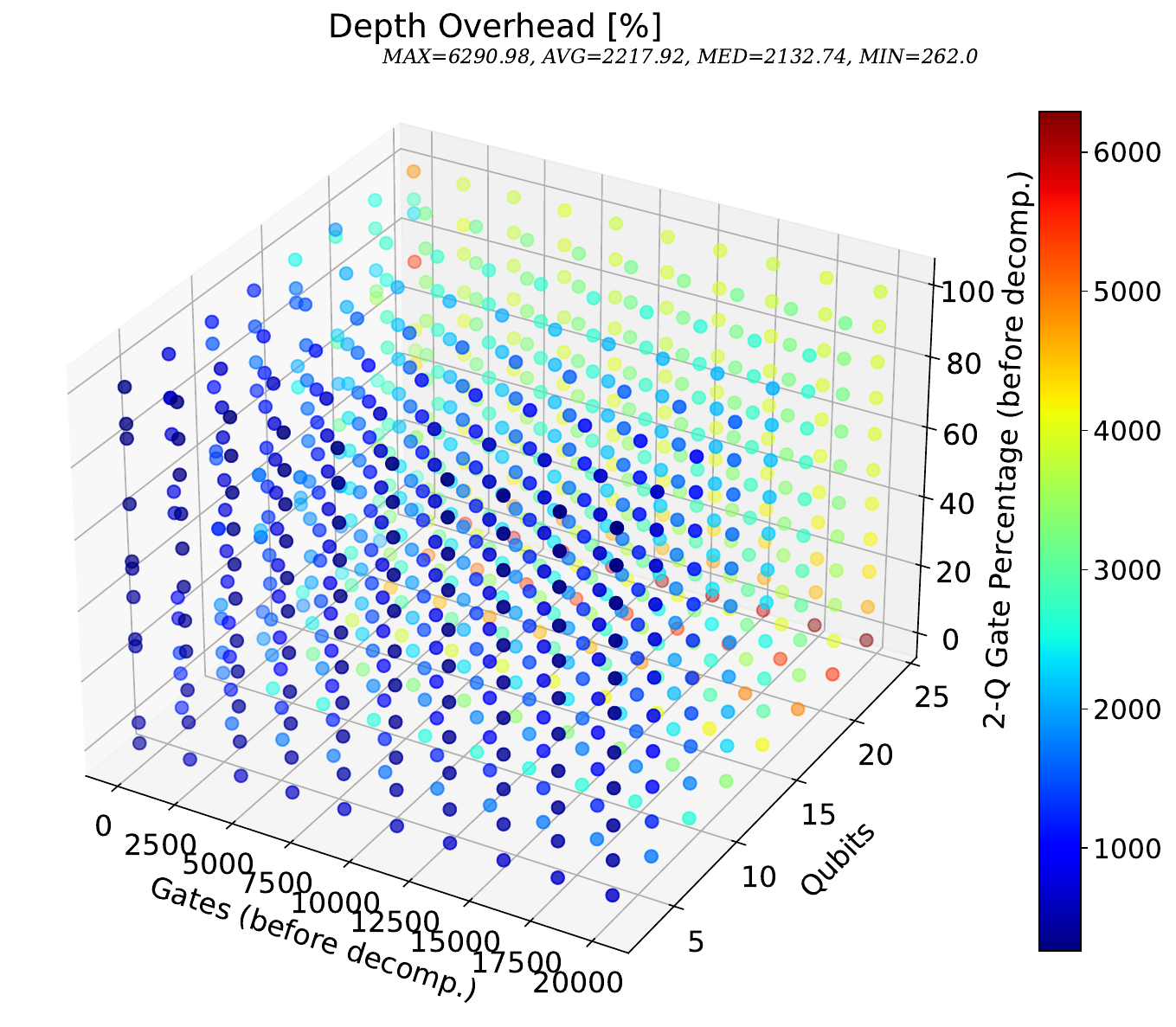}
         \caption{}
         \label{fig:depthrandom_small}
     \end{subfigure}
     \hfill
     \begin{subfigure}[b]{0.5\textwidth}
         \centering
         \includegraphics[width=\textwidth]{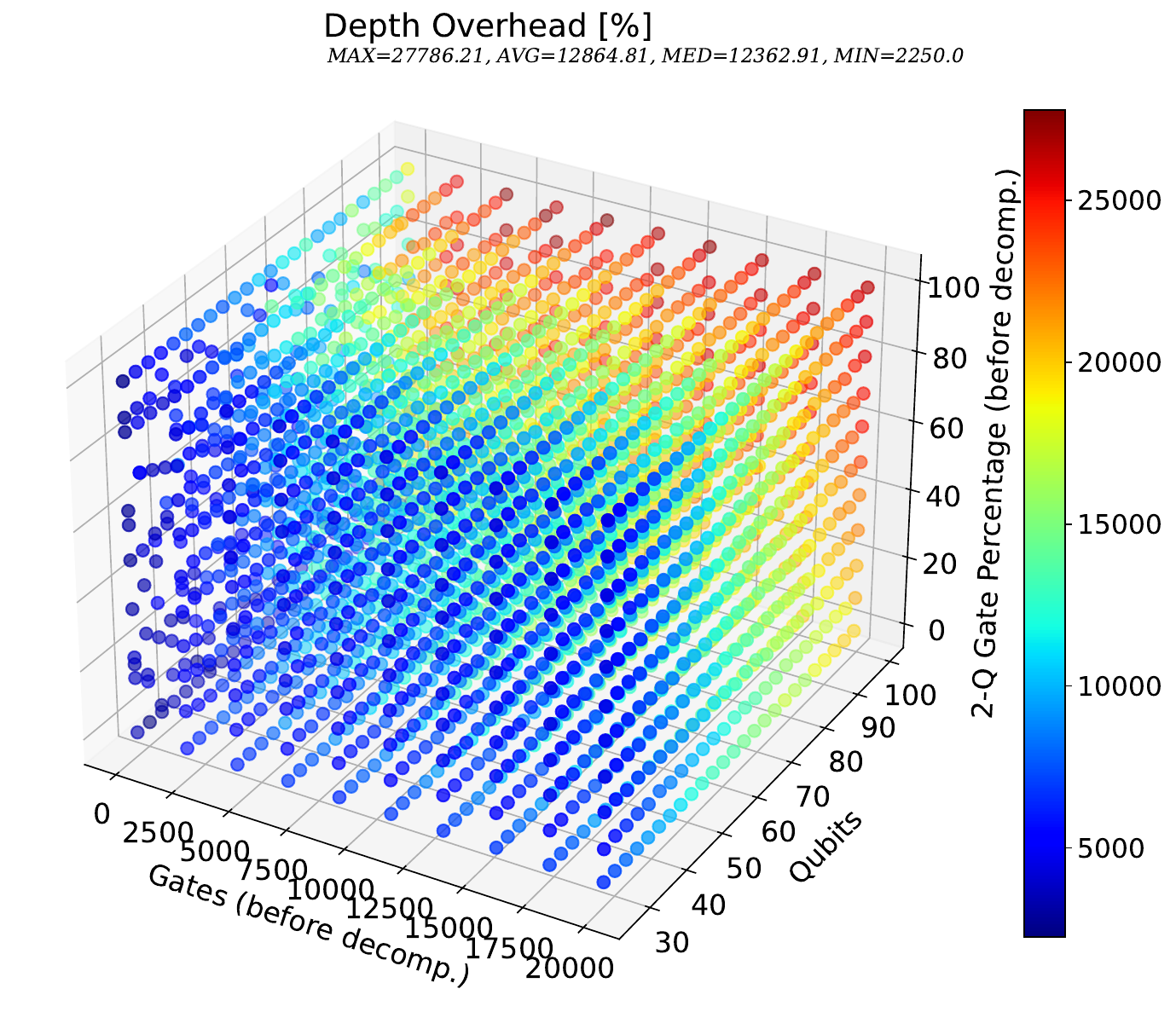}
            \caption{}
         \label{fig:depthrandom_medium}
     \end{subfigure}
    \caption{Resulting depth overhead when $3,630$ random uniform quantum algorithms are mapped onto the crossbar architecture. The three axes correspond to benchmark characteristics, namely, number of gates [50 - 20,000], number of qubits [3 - 99] (split into two subfigures), and two-qubit gate percentage [0\% – 100\%].}
    \label{fig:depth_random}
  
\end{figure}

\begin{figure}[htpb]
    \centering
    \includegraphics[width=0.5\textwidth]{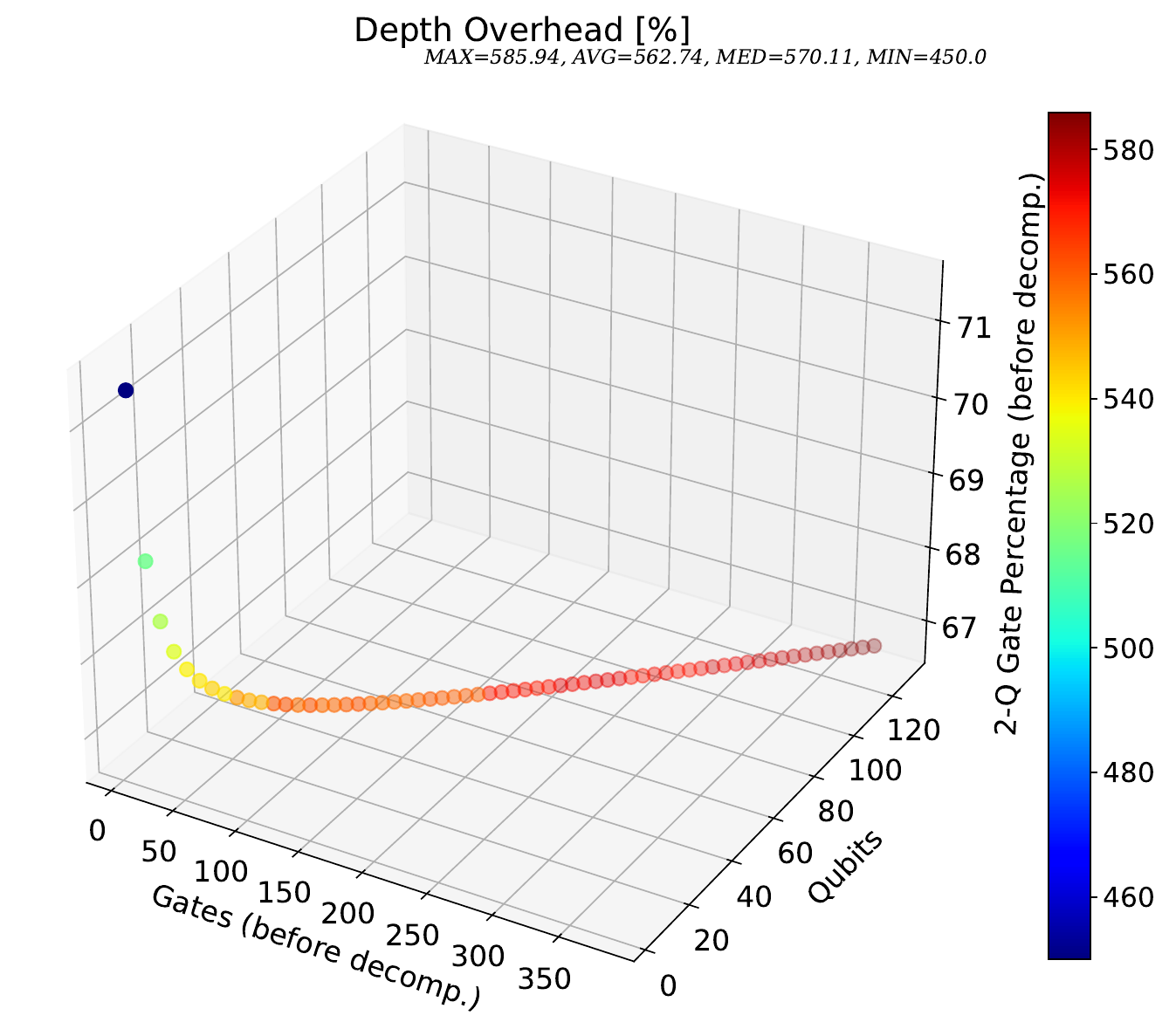}
    \caption{Resulting depth overhead when Cuccaro Adder from the Qlib library is mapped onto the crossbar architecture. The three axes correspond to benchmark characteristics, namely, number of gates [4 - 385], number of qubits [4 - 130] and two-qubit gate percentage [66.75 - 71.43].}
    \label{fig:depth_cuccaro_adder}
  
\end{figure}

\begin{figure}[htpb]

     \centering
     \begin{subfigure}[b]{0.5\textwidth}
         \centering
         \includegraphics[width=\textwidth]{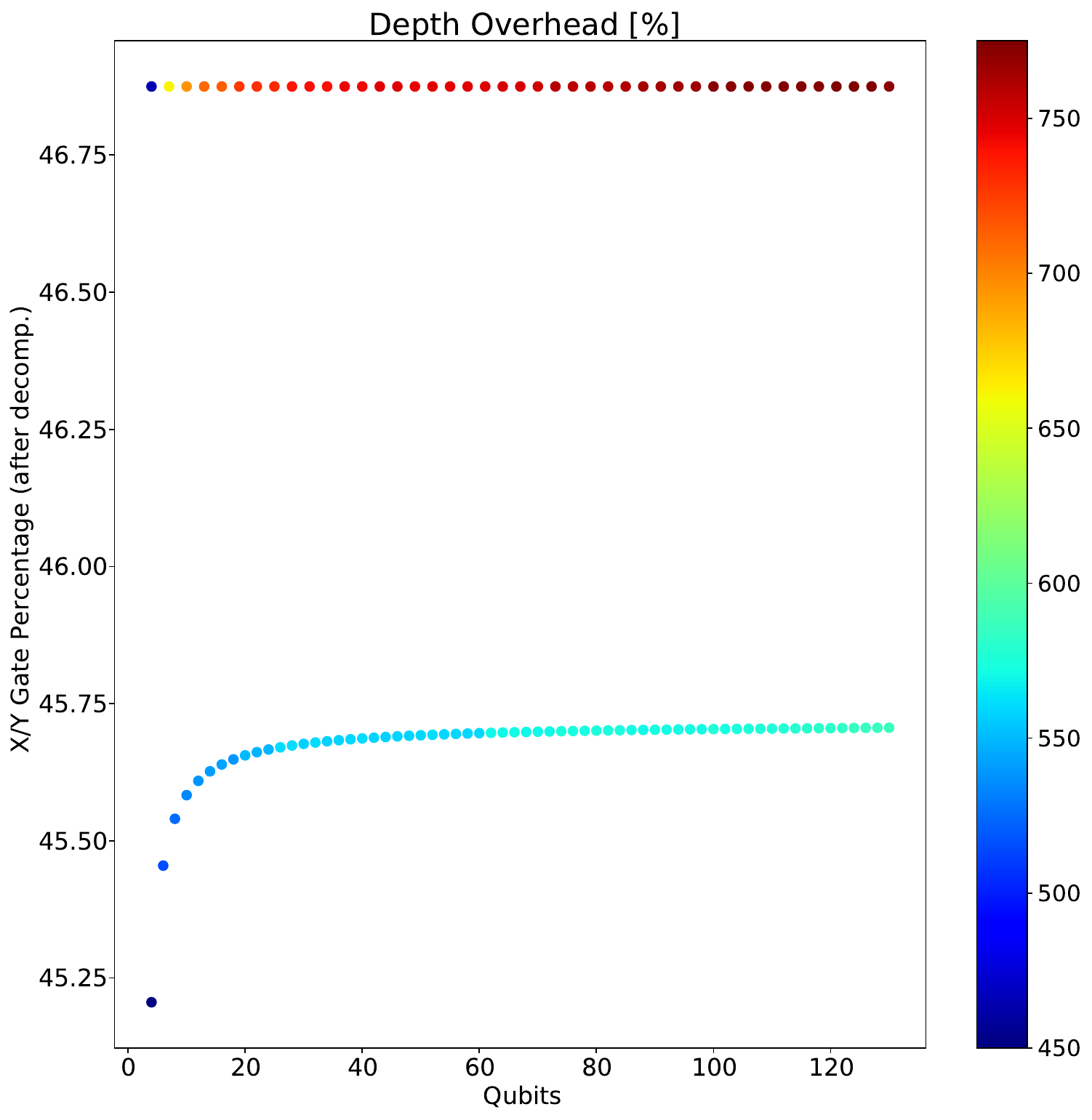}
         \caption{}
         \label{fig:depth_cuccaro_vbe_adder}
     \end{subfigure}
     \hfill
     \begin{subfigure}[b]{0.5\textwidth}
         \centering
         \includegraphics[width=\textwidth]{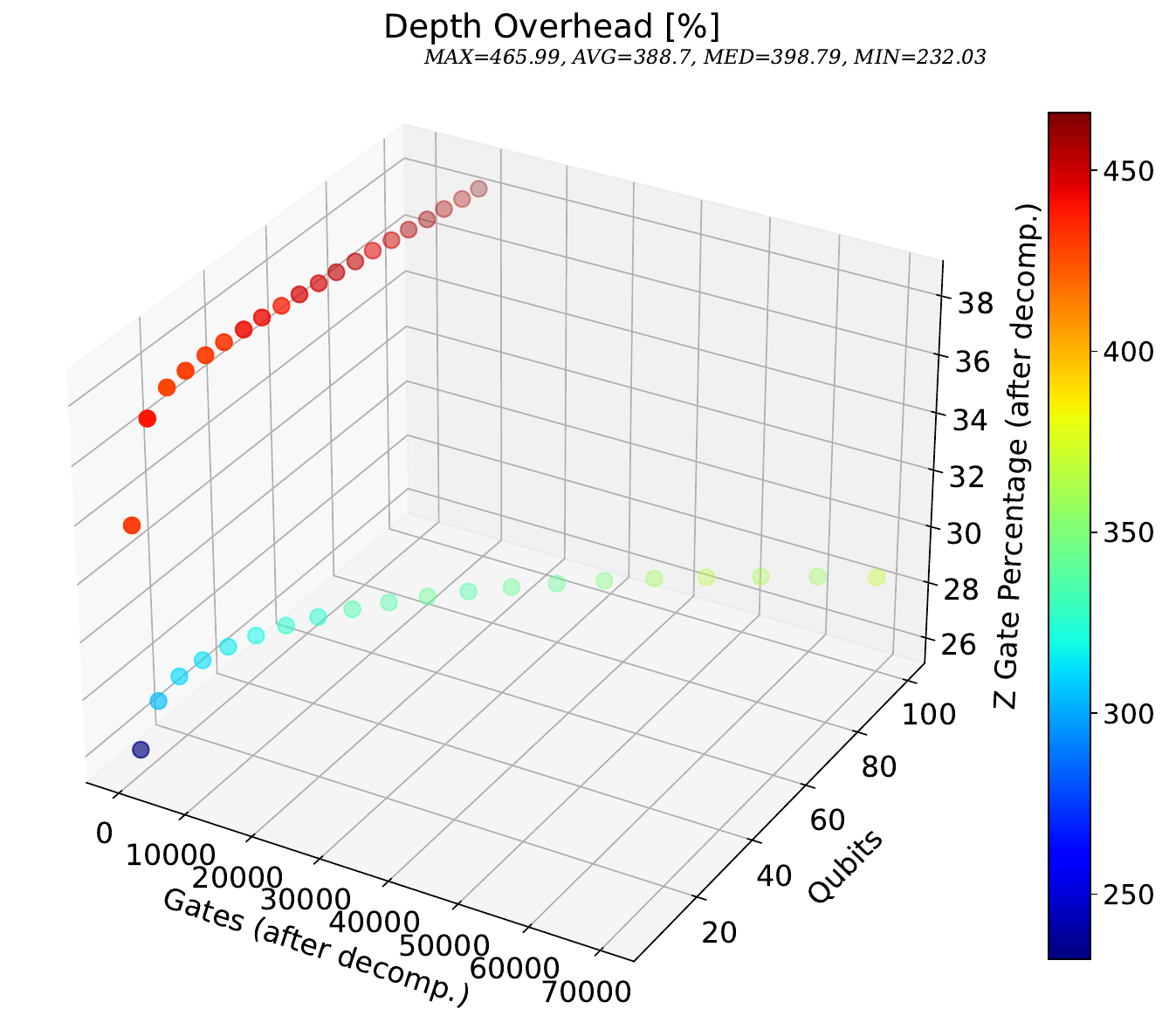}
            \caption{}
         \label{fig:depth3d_qft_grover_after}
     \end{subfigure}

    \caption{\textcolor{black}{(a) Resulting depth overhead when Cuccaro Adder (bottom line of data points) and Vbe Adder (top) from the Qlib library are mapped onto the crossbar architecture. The y-axis represents the $X$ or $Y$ gate percentage after decomposition, and the x-axis the number of qubits. (b) Resulting depth overhead when mapping Grover's (top line of data points) and QFT (bottom line of data points) quantum algorithms onto the crossbar architecture. The three axes correspond to benchmark characteristics after decomposition, namely, number of gates [52 - 20050], number of qubits [5 - 100], and Z gate percentage [25.97 - 38.29].}}
    \label{fig:depth_cuccaro_vbe_adder}   
\end{figure}

This time, we analyze in Fig. \ref{fig:depth_random},  the depth overhead when mapping onto the crossbar the same random uniform benchmark set as in Fig. \ref{fig:gaterandom}. It can be observed that the trend (colors) of the depth overhead changes for different ranges of number of qubits  as shown in the two subfigures. Knowing that the main source of depth overhead originates from $X$ or $Y$ gates  (at least $3$ additional cycles), we expect the depth overhead to become higher in lower regions of two-qubit gate percentage. That is observed in Fig. \ref{fig:depthrandom_small}, where the number of qubits goes up to 25. However, moving on to Fig. \ref{fig:depthrandom_medium}, we see that this trend changes. Now, due to the higher number of qubits, routing distances have increased, thus routing for two-qubit gates dominates the depth overhead. This is apparent by its increase (from blue to red color) as we go from lower qubit counts to higher qubit counts, and as we go from low to higher percentage of two-qubit gates. Finally, this fact is also apparent in the absolute values of depth overhead of the two subfigures. Note also that the number of gates has a slight influence on the depth overhead, but it is not as relevant as the other characteristics discussed above.

Moving on, Fig. \ref{fig:depth_cuccaro_adder} shows the depth overhead of a Cuccaro Adder when scaling it up from $4$ to $130$ qubits. In the range of $4$ to $20$ qubits, we observe an increase in depth overhead as the percentage of two-qubit gates decreases, which aligns with the remarks about the main source of depth overhead (i.e., the $X$ or $Y$ gates). Then, for an increasing number of qubits (from $20$ qubits on) and at an almost constant two-qubit gate percentage ($67\%$), the depth overhead increases at a slower rate. Here we conclude, once again, that two-qubit gate routing starts to dominate the depth overhead as routing distances become larger.

In most previous works, the amount of two-qubit gates is the main circuit characteristic to anticipate how much qubit routing will be needed for a specific quantum  algorithm and therefore the major and only source of gate/depth overhead. However, in the crossbar architecture, and potentially in other spin-qubit crossbar designs, single-qubit gates can also contribute to this overhead as discussed before. It is then important to have a closer look at the  $X$ or $Y$ rotation gate percentage and further analyze how it impacts the depth overhead. Additionally, after the gate decomposition step, the percentages and ratios between all gate types are changed. To illustrate this, imagine a quantum circuit  that originally consists of a low number of $CNOT$ gates and no $Z$ gates. After the decomposition to gates supported by the crossbar architecture, the percentage of $Z$ rotation gates will increase, and consequently, the two-qubit gate percentage will decrease, as  $CNOT$ gates are decomposed as $Ry(\frac{\pi}{2})$, two $\sqrt{SWAP}$, $S$, $S$\dag, $Ry(\frac{-\pi}{2})$. Thus, it is relevant to consider this gate percentage change in our analysis as ultimately the executable circuit will only consist of native gates. To summarize, as overhead comes from mapping different types of gates on the crossbar, individually distinguishing between them, in particular after decomposition, can increase the accuracy of our evaluations.

To illustrate the previous point, in Fig. \ref{fig:depth_cuccaro_vbe_adder} (a) we show the depth overhead of the Cuccaro Adder (upper dots) and the Vbe Adder (lower dots) with the same ranges as in Fig. \ref{fig:gatevbecucCOMP}. Note that the y-axis corresponds to the percentage of $X$ or $Y$ rotation gates after decomposition. From this new perspective, we clearly see their difference in actual (i.e., executed by the architecture) $X$ or $Y$ rotation gate percentage. On average the depth overhead of the Vbe adder is $196\%$ higher than the Cuccaro Adder for the same range of qubits. As explained before, the highest source of depth overhead comes from $X$ or $Y$ rotations gates, which explains the large depth overhead difference between those two algorithms. 

\textcolor{black}{In constant, in Fig. \ref{fig:depth_cuccaro_vbe_adder} (b), we show the depth overhead after decomposition of the same algorithms with the same ranges as in Fig. \ref{fig:gaterevlib} (b). This time, Grover's algorithm (top dots) shows on average $113\%$ higher gate overhead than of QFT algorithm (lower dots). Note here, these two algorithms are plotted in relation to their Z gate percentage in the z-axis. Thus, their performance variations can be partially attributed to their differences in Z gate percentage, though this cannot be a definitive explanation. As previously stated, drawing a direct comparison is less straightforward due to disparities in algorithm structure and differing rates of benchmark characteristics.}

\subsection{Insights from depth overhead analysis} \label{Insights from depth overhead analysis}

From the previous analysis, we can observe that trends can change based on the parameter ranges of benchmarks. This is because different sources of depth overhead contribute with different rates based on the number of qubits (i.e., crossbar size). More specifically, the overhead contribution resulting from mapping $X$/$Y$ gates was higher up to a certain number of qubits after which was exceeded by the contribution rate of two-qubit gates. We saw that exceeding a threshold of more than $20$ qubits increases the depth overhead at a steadier pace, which specifically favored scalability for Cuccaro Adder in Fig. \ref{fig:depth_cuccaro_adder} and \ref{fig:depth_cuccaro_vbe_adder}. It is expected, however, that with different algorithms, there will be different trends. With such observations, we stress the importance of distinguishing between all gate types and especially after decomposition to better understand the performance impact of mapping. With that knowledge, we can create better mapping techniques and/or make an informed selection of algorithms to execute. 

As mentioned before, the fact that gate overhead \textcolor{black}{and routing} can result from mapping single-qubit gates is unprecedented. Furthermore, we notice that mapping both, single- and two-qubit gates, requires additional shuttles and they produce the highest gate and depth overhead. Therefore, novel mapping techniques minimizing all qubit movements (shuttles) can increase performance substantially, such as the ones discussed in Sec. \ref{Insights from gate overhead analysis}. From an architectural point of view, since the shuttle operation is so relevant, there have to be as few operational constraints as possible when mapping them.

\textcolor{black}{Finally, the current SpinQ version does not parallelize gates that are shuttle-based. These are the resulting shuttle gates from the shuttle-based SWAP,  mapping of single-qubit gates, and the two shuttling operations to facilitate a Z rotation. An improved version can involve a constraint and conflict check for any shuttle-based type gate to reach the full parallelization potential of the second pass, without increasing the time complexity.}

\subsection{Estimated Success Probability} \label{Estimated Success Probability}
In this section, we will show how the success probability of an algorithm drops after mapping it to the crossbar architecture. Before we continue, we have to mention that even with operational fidelities as high as 99.99\% for single-qubit gates and shuttles (as suggested in \cite{li2018crossbar}) and 99.98\% for \(\sqrt{SWAP}\)s, the ESP drops drastically to $0$ in most algorithms with a high number of gates. For that reason, we only focused on the Bernstein-Vazirani algorithm as it has a low percentage of two-qubit gates (usually there are only one or two $CNOT$s), therefore errors are mostly introduced by single-qubit gates.

\begin{figure}[htpb]
    \centering
    \includegraphics[width=0.5\textwidth]{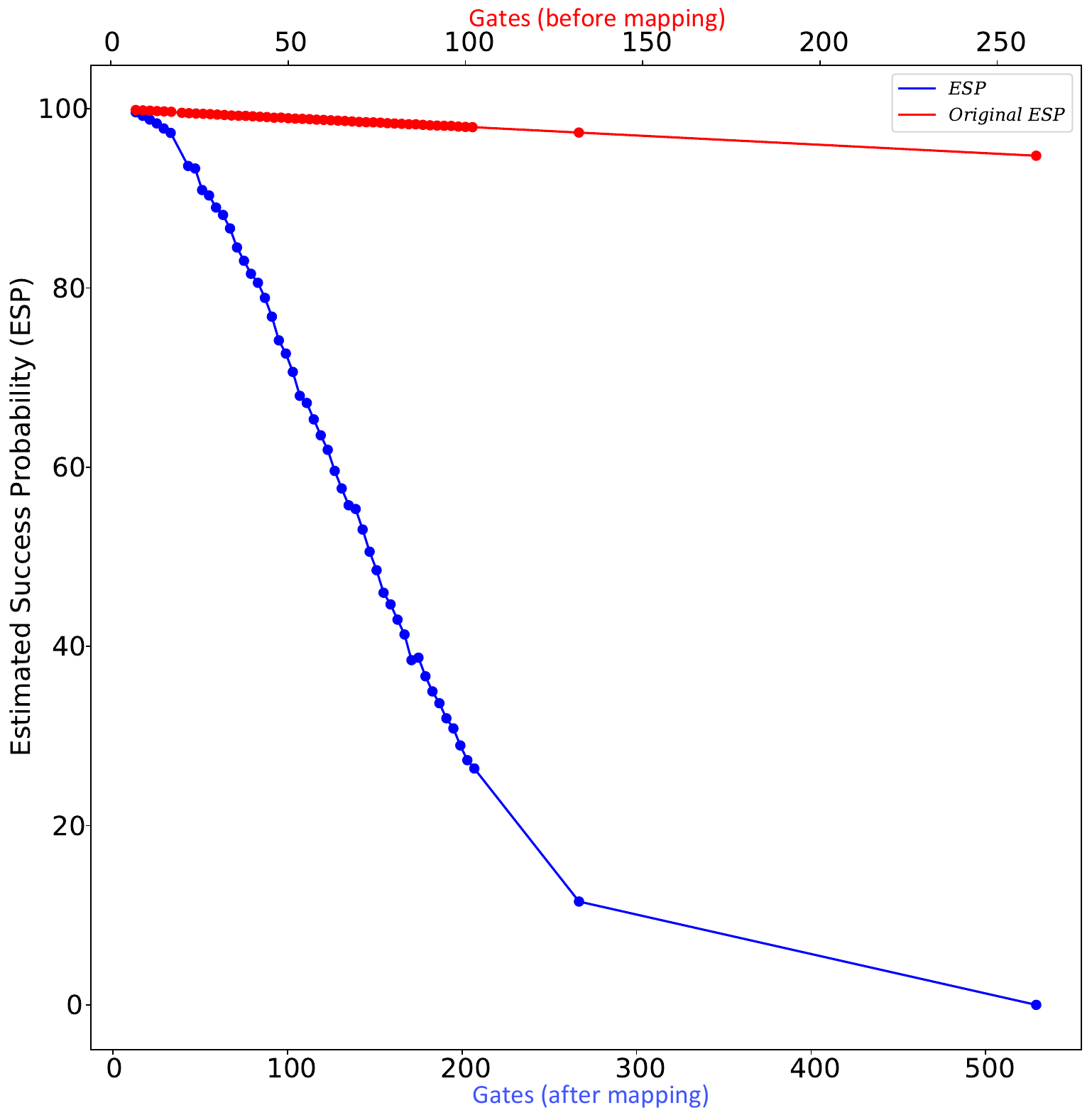}
    \caption{Estimated success probability (ESP) before and after mapping of Bernstein-Vazirani algorithm from $2$ to $129$ qubits.}
    \label{fig:ESP}      
\end{figure}

Fig. \ref{fig:ESP} shows the ESP of the Bernstein-Vazirani algorithm when scaling it from $2$ to $129$ qubits. The red line “Original ESP” refers to the ESP before mapping, and the blue line "ESP" refers to ESP after mapping. We observe a sharp ESP decrease approaching $10\%$ for $267$ gates after mapping with a slope rate of $-0.6$ which is caused by the increased number of gates. For $529$ gates after mapping we obtained a $0\%$ ESP.  Another reason for the ESP decrease is the semi-global single-qubit rotation; for each of the $X$ or $Y$ gates contained in the circuit after decomposition, all qubits in odd or even columns are rotated even the ones that are not targeted for rotation. This is further explained in Sec. \ref{Application of X or Y rotations on single qubits}.

\subsection{Insights from Estimated Success Probability analysis} \label{Insights from Estimated Success Probability analysis}

We observed a rapid decline in ESP in a minimally connected algorithm (mostly $X$ or $Y$ rotation gates), even though our equation did not include decoherence-induced errors \cite{helsen2018quantum,kharkov2022arline}. Although simple, equation \ref{eq:ESP_1}, is approximating a worse-case-scenario algorithm success rate. The main reason for this decrease is the resulting overhead when implementing single-qubit gates on specific qubits given the semi-global rotation scheme. Note that in this case, all qubits in either column parities are rotated thus each contributing to this ESP drop. Therefore, it is essential to determine which algorithms could take advantage of the semi-global control and/or develop architecture-specific mapping techniques to minimize the need for a scheme, as suggested in Sec. \ref{Insights from gate overhead analysis}.

There are other sources of noise on real NISQ quantum devices that impact algorithm execution. Fortunately, it is expected that processors will gradually become more robust with better fabrication tolerances and error-mitigation techniques that will enable quantum error correction protocols. It remains challenging, however, to accurately simulate errors in large-scale devices to derive the algorithm's success probability.

\subsection{Compilation time} \label{Compilation time}

\begin{figure}[htpb]
    \centering
    \includegraphics[width=0.5\textwidth]{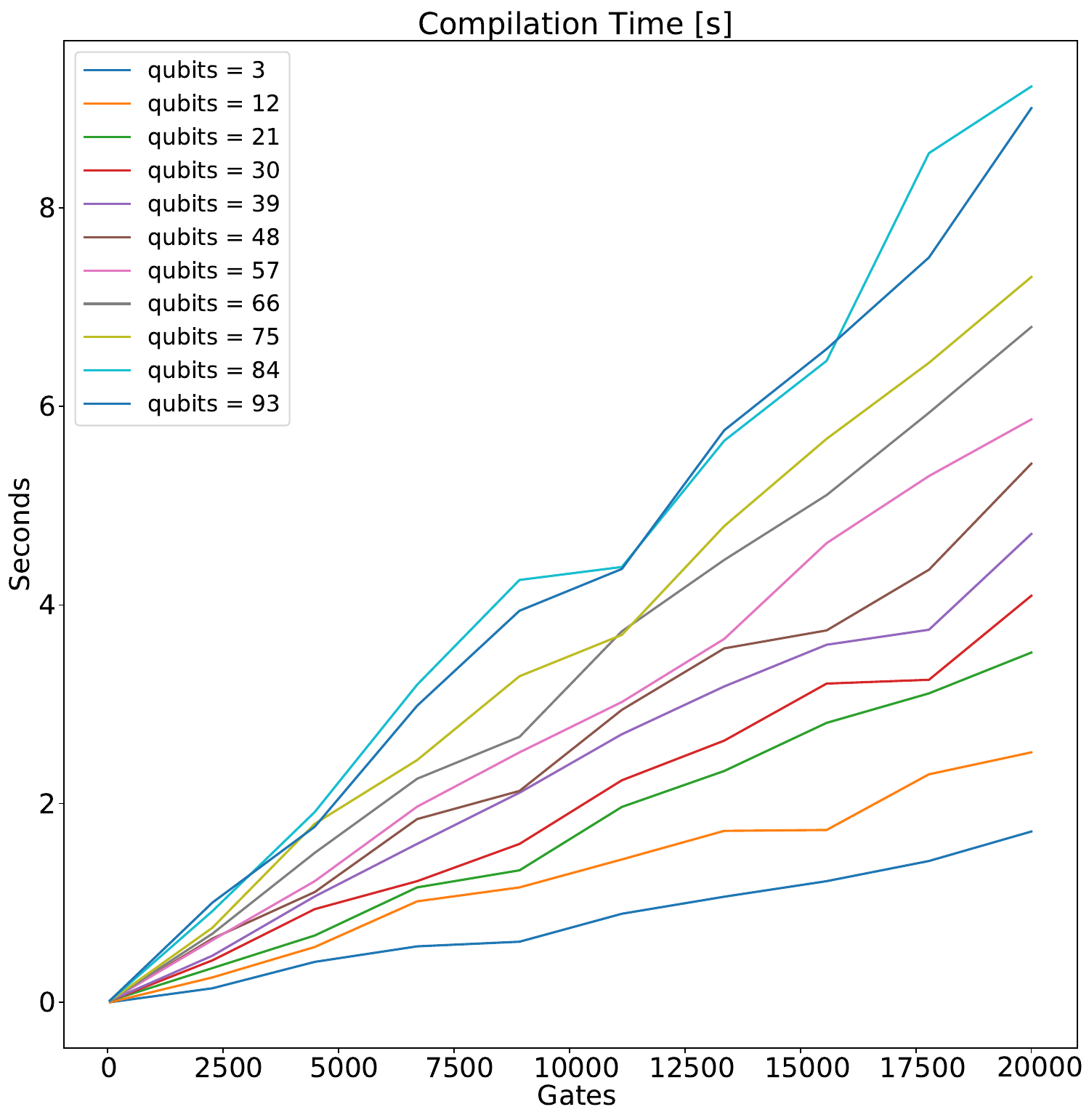}
    \caption{Compilation time when mapping random uniform algorithms with 50\% of two-qubit gates onto the crossbar architecture. We observe a linear relation that makes \textit{SpinQ} suitable for scalable crossbar architectures.}
    \label{fig:complilation_time}
     
\end{figure}

Finally, we measure the compilation time of our solution to evaluate its scalability. The compilation time of \textit{SpinQ Integrated Strategy} can be seen in Fig. \ref{fig:complilation_time} for a subset of the random uniform circuits that have been used in Fig. \ref{fig:gaterandom} and Fig. \ref{fig:depth_random}. This subset consists of circuits with only 50\% of two-qubit gates. With this subset we map the same number of gates for each gate type, thus all internal \textit{SpinQ} processes are weighted equally. We observe a linear increase (\textit{O(n)}) in compilation time in relation to the number of gates for each qubit count. This implies that our strategy is suited for scalable spin-qubit crossbar architectures. Improvements  \textcolor{black}{in compilation strategies of \textit{SpinQ}} can be directed towards reducing the slopes for each qubit count. This is further discussed in Sec. \ref{Integrated strategy improvements}.

\section{Discussion and future directions} \label{Discussion and future directions}

\subsection{Integrated strategy improvements} \label{Integrated strategy improvements}

\textcolor{black}{This is the first version of \textit{SpinQ} and the \textit{Integrated Strategy} is not utilized completely. More specifically, in this implementation, we are only parallelizing $X$ or $Y$ rotation gates and shuttles from the shuttle-based SWAP technique.}  However, there can be a few extensions to the \textit{Integrated Strategy} that can provide better performance (less overhead and higher ESP). These improvements can be divided into two categories: a) improvements that increase complexity marginally and b) improvements that will increase complexity substantially. It is important to make this differentiation because on large scale we have to consider the trade-off between complexity manifested as higher computation time as sizes increase and performance manifested as less overhead and higher ESP.

Improvements in category (a) \textcolor{black}{ are for the first and second pass of the \textit{Integrated Strategy} and will improve scheduling of gates that use the $QL$ lines (i.e., shuttling, two-qubit gates and Z rotations). Note that the \textit{Integrated Strategy} is made to support such an improvement within the two passes and it will not increase its time complexity. In particular, the first pass could parallelize ideally Z rotations together with one two-qubit gate, in lines 13 and 21 respectively. Then in the second pass in lines 35 and 37 a similar conflict checking process can be followed as line 37 when completing the Z rotation mapping and schedule them in the least cycles possible. Furthermore, two-qubit gate routing could be better parallelized in the second pass instead of the pairing presented in Fig. \ref{fig:shuttle_based_SWAP}. Two-qubit gate ideal parallelization in the first pass constitutes routing in the second pass which will require a new routing algorithm to handle multiple gates at the same time. Thus, such an improvement belongs to category (b).} Once again, each cycle remains dedicated to one gate type, therefore, fine-tuning pulse durations in real devices is still possible.

Moving on to the next category (b), it consists of all heuristic mapping algorithms (routing and initial placement) discussed in Sections \ref{Insights from gate overhead analysis}, \ref{Insights from depth overhead analysis} and \ref{Insights from Estimated Success Probability analysis}, which can be extended to other scalable spin-qubit architectures. This will enable complete parallelization of two-qubit gates and less routing for both, single- and two-qubit gates.

\textcolor{black}{Finally, in Sec. \ref{Measurement} an ideal measurement process has been considered for this work. However, it would increase the accuracy of our performance evaluation to include a more realistic readout process in \textit{SpinQ} as an additional step after the \textit{Integrated Strategy}. To do that there needs to be a protocol to move data and ancilla qubits in an efficient manner during and after algorithm execution --- a non-trivial task. An optimization algorithm will need to initialize ancilla qubits in the right places and time so they do not remain inactive for long periods. Then, optimize the measurement procedure to take as less steps and shuttling operations as possible. As a consequence, there will always be additional operational overhead and degradation of the algorithm's success during readout. Additionally, parallelization of qubit measurement is also a relatively unexplored topic and it highly depends on the ever-developing hardware implementations of spin-qubits. }

\subsection{Strategy Comparisons} 

\begin{table}[htpb]
\centering
\caption{Computational complexity comparison between compilation strategies for the crossbar architecture \cite{li2018crossbar}. With $n$ we denote the number of gates in a quantum circuit.}
\label{tab:comparison}
\begin{tabularx}{\columnwidth} { 
| >{\centering\arraybackslash}X 
| >{\centering\arraybackslash}X 
| >{\centering\arraybackslash}X 
| >{\centering\arraybackslash}X 
| >{\centering\arraybackslash}X | }

\hline
\textbf {Strategy} & \textbf{Complexity} \\
\hline
\textit{Backtrack} \cite{morais2019mapping} &	$O(n^3)$ \\
\hline
\textit{Suffer a side effect} \cite{morais2019mapping} &	$O(n^2 log(n))$	 \\
\hline
\textit{Avoid the deadlock} \cite{morais2019mapping} & $O(n)$  \\
\hline
\textit{Integrated} (ours) & $ O(n)$\\
\hline
\end{tabularx}
\end{table}

As we discussed in Sec. \ref{Mapping challenges of a crossbar architecture}, the crossbar architecture comes with constraints that prevent full parallelization of quantum instructions. The crossbar, however, may reach two types of conflicts (i.e., unwanted interactions or blocked paths), even if all constraints are respected. For that reason, there must be some kind of compilation strategy between the scheduler and the router to prevent conflicts. In this work, we have implemented the \textit{Integrated strategy} which is different from the three strategies suggested in \cite{morais2019mapping}. Table \ref{tab:comparison} compares the computational complexity of these three strategies with our own. The \textit{Backtrack} strategy suggested in \cite{morais2019mapping} avoids conflicts by trying alternative scheduling combinations. If after repeating this process the scheduler has backtracked to the first instruction of the cycle, meaning no more scheduling combinations, a new routing path is generated by the routing algorithm and the scheduling is repeated. This strategy can be quite complex as the worst case scenario can un-route and un-schedule all the gates going back to a completely un-mapped circuit until a conflict-free mapping is found. An improved version of this strategy, called \textit{Suffer a side effect}, is a special case of the former and it is only preferred whenever a corresponding conflict can be corrected and if the correction is less costly than exclusively following the "backtracking" strategy. The final strategy, and the one implemented in \cite{morais2019mapping}, is called \textit{Avoid the deadlock}. This strategy, similar to our \textit{Integrated strategy}, is trying to avoid conflicts by parallelizing only $X$ or $Y$ gates. In this way, \(\sqrt{SWAP}\)s and shuttle operations can not cause a conflict. However, in this strategy there is no synergy between the routing and scheduling stages as our \textit{Integrated strategy} has, therefore there is little flexibility for improvements and performance can not be easily improved while maintaining the same complexity. Our strategy is able to maintain the same $O(n)$ complexity even after improvements, particularly type (a) improvements referred to in Sec. \ref{Integrated strategy improvements}

\subsection{General discussion} 
When developing novel mapping techniques for scalable quantum computing architectures such as the si-spin crossbar two main factors have to be considered: \textit{scalability} and \textit{adaptability}. As spin-qubit fabrication capabilities are improving, new architectural designs with potentially higher qubit counts will be explored. Therefore, from a computation/compilation time point of view, mapping techniques should be as scalable as the underlying technology. Practically, this implies that highly sophisticated and more complex mapping techniques might be excellent for a particular architecture and up to a certain number of qubits, but could be impractical for more qubits or even unusable for another architecture. In addition, as we are slowly exiting the NISQ era, quantum technologies will become more robust, especially with the use of  quantum error correction techniques. By that time, optimizing mapping techniques for specific hardware and/or algorithm might not be as relevant as today, but rather how fast and adaptable these techniques are to a plethora of quantum algorithms and increased number of qubits.

\section{Conclusion} \label{Conclusion}

Different quantum circuit mapping techniques have been developed to deal with the limitations that current quantum hardware presents and are being consistently improved to expand its computational capabilities by getting better and better algorithm success rates. The most advanced mapping methods focus on ion-trap and superconducting devices due to their `maturity' compared with other quantum technologies. However, spin-qubit-based processors have a great potential to scale rapidly and the first 2D crossbar architectures have been recently demonstrated.  In this work, we focused on the quantum circuit mapping challenges of the newly emerging spin qubit technology for which highly-specialized mapping techniques are needed to take advantage of its operational abilities. Specifically, we used the crossbar architecture as a stepping stone to explore novel mapping solutions while focusing on scalability. The crossbar architecture adopts a shared-control scheme, thus making it a great candidate to tackle the interconnect bottleneck. On that note, we have developed \textit{SpinQ}, the first native compilation framework for spin-qubit architecture which we used to analyze the performance of synthetic and real quantum algorithms on the crossbar architecture. Through our analysis, we tried to inspire novel algorithm- and hardware-specific mapping techniques that can increase the performance while taking into account compilation scalability. We also emphasized the importance of characterizing benchmarks before and after decomposition together with their Quantum Interaction Graph (QIG) structure to better evaluate results. We plan to make \textit{SpinQ} publicly available in the future.

\section{Acknownledgement}
This work is part of the research program OTP with project number 16278, which is (partly) financed by the Netherlands Organisation for Scientific Research (NWO). This work has also been partially supported by the Spanish Ministerio de Ciencia e Innovación, European ERDF under grant PID2021-123627OB-C51 (CGA). We thank Menno Veldhorst and Hans van Someren for their fruitful discussions.
\section{Acknownledgement}

\bibliographystyle{ACM-Reference-Format}
\bibliography{bibliography}

\end{document}